\documentclass[aip,floatfix,amsmath,amssymb,reprint,superscriptaddress]{revtex4-2}

\usepackage{graphicx}
\usepackage{dcolumn}
\usepackage{bm}
\usepackage[colorlinks=True,allcolors=blue]{hyperref}
\usepackage{newtxtext,newtxmath}
\usepackage{microtype}
\usepackage{orcidlink}
\usepackage{lipsum}
\usepackage{amsmath,amssymb}
\usepackage{pdfpages}
\usepackage{sidecap}
\usepackage[detect-all]{siunitx}
\newcommand{\ecfp}{Edinburgh Complex Fluids Partnership, School of Physics and Astronomy, The University of Edinburgh, James Clerk Maxwell Building, Peter Guthrie Tait Road, Edinburgh EH9 3FD, United Kingdom}

\makeatletter 
\AtBeginDocument{\let\LS@rot\@undefined}
\makeatother
\makeatletter
\renewcommand\@make@capt@title[2]{%
 \@ifx@empty\float@link{\@firstofone}{\expandafter\href\expandafter{\float@link}}%
  {\textbf{#1}}\@caption@fignum@sep#2\quad
}%
\makeatother

\newcommand{\rmd}{\mathrm{d}}
\newcommand{\m}{m}
\newcommand{\B}{B}
\newcommand{\hg}{h_g}
\newcommand{\weff}{w_\mathrm{eff}}
\newcommand{\etaf}{\eta_\mathrm{eff}}
\newcommand{\rim}{\dot\gamma_w}
\newcommand{\vin}{v_i}
\newcommand{\hin}{h_i}
\newcommand{\mmpm}{\milli\metre\per\minute}
\newcommand{\Figref}[1]{Fig.~\ref{#1}}
\newcommand{\partFigref}[2]{Fig.~\hyperref[#1]{\ref*{#1}#2}}
\newcommand{\simplepartFigref}[2]{part~\hyperref[#1]{#2}}
\newcommand{\Mopt}{M_{\max}^{\rm opt}}
\newcommand{\eg}{\textit{e.g.}}
\newcommand{\ie}{\textit{i.e.}}
\newcommand{\cf}{\textit{cf.}}

\begin{document}

\title[Non-Newtonian fluid impact protection of laminates]{Optimising non-Newtonian fluids for impact protection of laminates}

\author{James A.\ Richards\,\orcidlink{0000-0002-2775-6807}}
 \email{james.a.richards@ed.ac.uk}
\author{Daniel J.\ M.\ Hodgson\,\orcidlink{0000-0002-9969-254X}}
\author{Rory E.\ O'Neill\,\orcidlink{0000-0002-4901-7087}}
    \affiliation{\ecfp}
\author{Michael E. DeRosa}
    \affiliation{Corning Research and Development Corporation, Corning, New York 14831, USA}
\author{Wilson C.\ K. Poon\,\orcidlink{0000-0003-0760-7940}}
    \affiliation{\ecfp}

\begin{abstract}
    Non-Newtonian fluids can be used for the protection of flexible laminates. Understanding the coupling between the flow of the protecting fluid and the deformation of the protected solids is necessary in order to optimise this functionality. We present a scaling analysis of the problem based on a single coupling variable, the effective width of a squeeze flow between flat rigid plates, and predict that impact protection for laminates is optimised by using shear-thinning, and not shear-thickening, fluids. The prediction is verified experimentally by measuring the velocity and pressure in impact experiments. Our scaling analysis should be generically applicable for non-Newtonian fluid-solid interactions in diverse applications.
\end{abstract}


\maketitle
\noindent\fbox{%
    \parbox{0.975\columnwidth}{%
\textsf{Complex fluids that alter their mechanical response as the applied forces change enable smart materials. A prime example is flexible body armour infused with a shear thickening suspension that hardens on impact. During impact there is a complex interplay between solid deformation and fluid flow that complicates predictive design. We construct and experimentally validate a theoretical model for a fluid-solid laminate that describes display glass applications, such as in smartphones. Strikingly, we find that, now, sandwiching a fluid that becomes less viscous during impact between a top and a bottom layer protects both against impact. Our approach establishes new design principles for smart fluid-solid composites.}}}

\vspace{0.5em}

Woven fabrics impregnated with a shear-thickening colloidal fluid, whose viscosity increases suddenly at a critical shear rate, can function as body armour~\cite{lee2003ballistic}. Perhaps surprisingly, the shear-thickening fluid does not provide protection in body armour because of their bulk rheology that allows, for example `running on cornstarch'~\cite{peters2016direct}. Instead, the fluid increases inter-fibre friction and so prevents fibres from being pulled apart~\cite{mawkhlieng2020review}, so that they form a rigid layer to spread impact and protect the material underneath. 

Partly inspired by this application, there is growing interest in smart materials that incorporate various non-Newtonian fluids in solid structures~\cite{cai2021impact,caglayan2020impact,fischer2006dynamic,pinto2017design,myronidis2022polyborosiloxane}. In particular, in direct analogy with body armours, it is envisaged that including shear-thickening fluids in laminates may provide impact protection. However, analysing the impact response of fluid-solid composites is challenging even in the case of Newtonian fluids~\cite{hou2012numerical}. Deformation of the solid drives fluid flow, which then generates a pressure, which in turn changes the solid deformation, creating  feedback. For a non-Newtonian fluid, such fluid-solid interaction is even more challenging, because the fluid property changes as the flow develops throughout impact, and analyses to date are limited, \eg, to blood flow~\cite{janela20103d,anand2019non,zhu2017deformable}, or stationary process such as blade coating~\cite{krapez2022spreading}.

We consider fluid-solid interactions in a laminate consisting of a non-Newtonian fluid sandwiched between a flexible sheet above and a rigid base below, which is a model for various real-life applications, \eg, a display in which the base layer is an LCD panel and the top layer is a piece of glass, both of which must be protected from concentrated impacts at $\lesssim\mathcal{O}(\SI{10}{\metre\per\second})$. The physics differs from that in shear thickening body armour. The requirement here is to protect {\it both} solid layers, while body armour is optimised for the protection of the single lower layer.

We perform a scaling analysis of the coupling between fluid flow, rheology and solid deformation in our geometry based on the idea of an `effective squeeze flow width', and verify our analysis using controlled-velocity impact experiments. We find that the effective squeeze flow width varies weakly throughout the impact, so that the process can be approximated as a simple rigid squeeze flow. From this we find, surprisingly, that shear thinning, not thickening, is optimal for protection.

\section*{Results}
\subsection*{Modelling}

\begin{figure}[t]
    \begin{center}
    \includegraphics[width=\columnwidth]{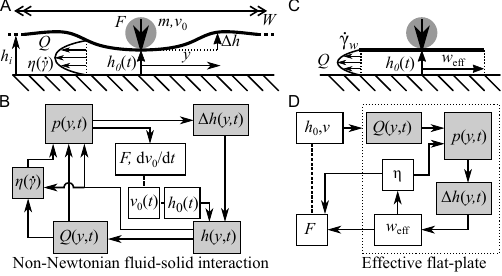}
    \end{center}
    \caption{Non-Newtonian fluid-solid interaction. \textit{(A)} Diagram of point impact on simplified laminate geometry. \textit{(B)}~Schematic of full coupling between fluid rheology, fluid flow and glass deformation. \textit{(C)}~Diagram of simplified effective plate. \textit{(D)}~Schematic of simplified closure with single effective plate width variable.}
    \label{fig:fsi}
\end{figure}

Using a quasi-2D setup, we analyse the downward impact of a point mass $\m$ at the origin, $y = 0$, with speed $v$ on a flexible plate initially at height $h_i$ parallel to a rigid bottom plate, with the gap filled by a fluid,~\partFigref{fig:fsi}{A}. The width of the plate $W\gg h_i$, and breadth of the plate (perpendicular to the page) $L\gg h_i$. The upper plate is pushed down, leaving a gap $h_0(t)$ at the impact point, and bending deformation $\Delta h(y,t)$ upwards. The net motion causes a fluid flow, $Q$. If the impact velocity is significantly sub-sonic,\ie\ $v_0 \ll \mathcal{O}(\SI{1000}{\metre\per\second})$ for most solids and liquids, then incompressibility and mass conservation require 
\begin{equation}
    \frac{\partial}{\partial t}\left[ h_0(t) + \Delta h (y,t) \right]=-\frac{\partial Q}{\partial y},
    \label{eq:volume}
\end{equation}
with $y$ the distance from the impact. The  pressure gradient associated with the impact-driven flow is given by
\begin{equation}
    \frac{\rmd p}{\rmd y} = - \frac{12\eta Q}{h^3}\,,
    \label{eq:Poiseuille}
\end{equation}
where the fluid viscosity $\eta$ is constant for a Newtonian fluid.
The pressure, $p(y,t)$, which satisfies $p(y \! \rightarrow \! \infty)=0$, pushes back on the impacting mass $\m$,
\begin{equation}
    \m\frac{\rmd v}{\rmd t} = \m\frac{\rmd^2 h_0(t)}{\rmd t^2} = - L\int_{\infty}^{+\infty}\!\rmd y\, p(y,t)\,,
    \label{eq:impacter}
\end{equation}
and bends the flexible layer, which has thickness $\hg$ and rigidity $B = EL\hg^3/12$ (with $E$ its Young's modulus). The shape of the layer follows the Euler-Bernoulli equation~\cite{bauchau2009euler},
\begin{equation}
    \frac{B}{L} \frac{\partial^4 \Delta h}{\partial y^4} = p(y,t),
    \label{eq:beam}
\end{equation}
where we have neglected the laminate mass as $\ll m$. Self consistency requires that \eqref{eq:beam} solves to give small plate deflection, so that flow is essentially along $y$, as is assumed in the `lubrication approximation'~\cite{hamrock2004fundamentals}, \eqref{eq:Poiseuille}.

The coupled integro-differential equations, Eqs.~(\ref{eq:volume}--\ref{eq:impacter}) need to be supplemented by a form for the rate-dependent viscosity, $\eta(\dot\gamma)$, if the fluid is non-Newtonian. The complex feedback between quantities, \partFigref{fig:fsi}{B}, means that finite element or immersed boundary numerical methods are needed to solve specific fluid-solid interaction problems for Newtonian~\cite{hou2012numerical} and non-Newtonian fluids~\cite{zhu2019ib,xu2016numerical}; but such solutions offer little physical insight into fluid-solid interactions, for which we turn to a different approach.

\subsubsection*{Simplified closure}

To analyse the fluid-solid interactions in our geometry, note first that since the pressure gradient $\partial_y p \propto h^{-3}$, we need only consider the region around the impact where deformation is small, $\Delta h \lesssim h_0$.\footnote{Initial contact is not accurately described, but for large deformations ($h_0 \to 0$) this can be neglected.} Within this region the surface is only weakly curved, and a calculation of the shear rate shows that it is adequate to treat it as a flat surface, $h(y)\approx h_0$ (SI Appendix, Fig.~S1). We therefore define an effective flat plate width, $\weff$, such that the pressure created by a rigid plate squeeze flow bends the flexible plate by $\Delta h = h_0$ at $y =\weff$. The squeeze flow for $|y|\leq w_{\rm eff} \ll W$ is solved analytically~\cite{gibson1998squeeze}, but we neglect fluid flow and deformation outside ($|y| > w_{\rm eff}$), \partFigref{fig:fsi}{C--D}. Within this local approximation, boundary conditions can be neglected as volume conservation will be ensured by, \eg, the surface being pushed up further away from the impact zone.

We use a scaling analysis to determine $\weff$, which is not known \textit{a priori}. The flux created by the rigid-plate squeeze flow $Q \simeq v\weff$ gives $\partial_y p \simeq 12\eta v \weff /h_0^3$ and $p \simeq 12\weff^2 \eta v /h_0^3$. Equation (\ref{eq:beam}) implies that the the deflection $\Delta h \simeq p \weff^4 \times L/B$. Self consistency demands that this $\Delta h \approx h_0$, which combines with $p$ to give
\begin{equation}
    \weff \simeq \left( \frac{B h_0^4}{12\eta v L} \right)^{\frac{1}{6}}, \frac{F}{L} \simeq p\weff \simeq \frac{12 \eta v \weff^3}{h_0^3} = \left(\frac{12\eta v \B}{Lh_0^2}\right)^{\frac{1}{2}}.
    \label{eq:FWnewt}
\end{equation}
While higher $\eta$, faster $v$ and narrower $h_0$ bend the plate more strongly and reduce $\weff$, the dependence is weak. The somewhat unusual $\frac{1}{6}$ exponent is traceable to the dependence of plate deflection on $\weff^6$.\footnote{Larger $\weff$ increases $Q$ and $\partial_{y}p \propto \weff$, such that $p \propto \weff^2$ and $F \propto \weff^3$. The bending moment in the plate $\propto \weff^4$, the angular deflection $\propto \weff^5$ and, ultimately, $\Delta h \propto \weff^6$.} The nearly-constant $\weff$ means that the dynamics can be thought of as a modified fixed width squeeze flow that scales approximately as $h_0^{-3}$.

To capture the lowest order effects of a rate-dependent viscosity, $\eta = \eta(\dot\gamma)$, in non-Newtonian fluids, a further approximation is made. We take the fluid to be an effectively Newtonian with a single viscosity, $\etaf = \eta(\rim)$, where $\rim$ is the shear rate at the edge of the effective plate  ($y = \weff$) for a fluid of this viscosity. This again ensures self-consistency; it also recalls the use of the rim shear rate in calculating the viscosity in parallel-plate rheometry~\cite{macosko1994rheology}.

We use a power-law model, $\etaf = K\rim^{n-1}$, to explore the effect of thinning ($n<1$) and thickening ($n>1$) on impact protection. Now, \eqref{eq:FWnewt} becomes (see SI Appendix)
\begin{equation}
    \begin{split}
    \weff \propto \left(\frac{\B h_0^{4}}{12KvL}\right)^{\frac{1}{6}}\left ( \frac{(6v)^5\B}{2h_0^8KL} \right)^\frac{1-n}{6(n+5)}\\
    \mathrm{and}~\frac{F}{L} \propto \frac{\sqrt{12K v \B/L}}{h_0} \left(\frac{(6v)^5\B}{2h_0^8KL} \right)^\frac{n-1}{2(n+5)}\!\!,
    \label{eq:wFsolved}
    \end{split}
\end{equation}
which reduce to Newtonian results, \eqref{eq:FWnewt}, for $K = \eta$ and $n = 1$. Equation~(\ref{eq:wFsolved}) gives the force per unit length in terms of $(B/L, K, n)$ and a single dynamical variable $h_0(t)$ with its derivative $\dot h_0 = v$; this then allows us to understand how a flexible solid-fluid laminate may be protected against impact.

\subsubsection*{Numerical solutions}

After impact, a time-dependent bending moment $M(t) = F(t) \weff(t)$ develops, which flexes the upper plate, \eqref{eq:beam}. Large flexure can lead to breakage when $M$ exceeds a critical bending moment, $M^*$. Protection requires minimising the maximum, $M_{\max}<M^*$, \eg, for a given geometry through fluid optimisation.

\onecolumngrid

\begin{SCfigure}[\sidecaptionrelwidth][t]
\includegraphics{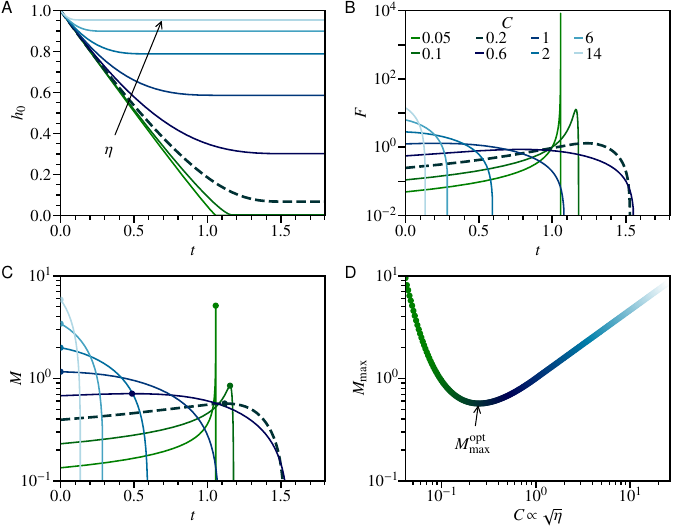}
    \caption{Predicted response to impact for a Newtonian fluid laminate with varying viscosity. \textit{(A)}~Changing gap, $h_0(t)$, normalising length $\hin$ and time $\hin/\vin$. Lines: green to light blue with increasing viscosity, $\eta$, setting impact parameter, $C = (12 \eta BL)^{0.5}/mv_{i}^{3/2}$, see legend in B. Bold dashed line for $C=0.2$ at optimum viscosity, see D. \textit{(B)}~Impact force, $F$, normalised by $(L\eta \vin B)^{0.5}/\hin$. \textit{(C)}~Bending moment, $M(t) = F w_{\rm eff}$, with effective plate width $w_{\rm eff}$ normalised by $(B\hin^4/L\vin \eta)^{1/6}$ and peak $M(t)=M_{\max}$ (circle). \textit{(D)}~Peak bending moment, $M_{\max}$ vs $C$.}
    \label{fig:newtonian}
\end{SCfigure}
\twocolumngrid

A Newtonian-fluid laminate with initial gap $\hin$ impacted by mass $m$ at initial downward speed $\vin$ obeys from \eqref{eq:impacter}
\begin{equation}
    \frac{\rmd^2 h_0}{\rmd t^2} = - \frac{C}{h_0} \left |\frac{\rmd h_0}{\rmd t} \right|^{\frac{1}{2}},~C=\sqrt{\frac{12 \eta BL}{m^{2}\vin ^{3}}}\,.
    \label{eq:NewtonianImpact}
\end{equation}
The gap and time have been normalised by $\hin$ and $\hin/\vin$, giving a single dimensionless `impact parameter', $C$, which captures the ratio of viscous dissipation, $F(h_i) \times h_i \propto \sqrt{v_i}$, to kinetic energy, $\propto v_i^2$. We solve for $h_0(t)$ numerically (using SciPy v1.10.1 \texttt{integrate.odeint}) for various $C \propto \sqrt{\eta}$, \Figref{fig:newtonian}.

At large $\eta$ ($C = 14,$ 6, 2), the impact is rapidly stopped and the gap hardly drops, $h_0(t\rightarrow \infty) \lesssim 1$, \partFigref{fig:newtonian}{A}~[light blue lines, see legend in \simplepartFigref{fig:newtonian}{B}]. This causes a large initial force, $F(0)$, \simplepartFigref{fig:newtonian}{B}, and maximum bending moment $M_{\max} = M(0)$ that grows with $\eta$, \simplepartFigref{fig:newtonian}{C}~(circle); however, both $F(t)$ and $M(t)$ drop rapidly. 
At intermediate $\eta$ ($C = 1$, 0.6), $h_0$ decreases noticeably before stabilising, while $F(0)$ and $M(0)$ both drop, but $F(t)$ and $M(t)$ stay constant for longer before dropping rapidly. At the smallest $\eta$ ($C = 0.1, 0.05$), the impact is not slowed and $h_0 \rightarrow 0$, giving a sharp peak in $F(t)$, \simplepartFigref{fig:newtonian}{B}, and in $M(t)$ (as $\weff$ changes sub-linearly with $h_0$) that now grows as $\eta \rightarrow 0$, \simplepartFigref{fig:newtonian}{C}. 

At some optimal $C\approx 0.2$, $M_{\rm max}$ is minimised at $\Mopt$, \simplepartFigref{fig:newtonian}{D}. The impact is absorbed over the whole gap with a near-constant $v = \dot h_0$, but eventually slows before $F$ diverges. As $\weff$ is weakly dependent on $h_0$, reducing the divergence in $F$ directly gives a flatter $M(t)$. This, however, still peaks as the gap narrows, \simplepartFigref{fig:newtonian}{C}~[bold dashed line], increasing 50\% from $t = 0$ before dropping rapidly to zero. To obtain a minimum $M_{\rm max}$ with a flat $M(t)$ profile, we turn to non-Newtonian fluids. 

\begin{figure}[t]
    \centering
    \includegraphics[width=\columnwidth]{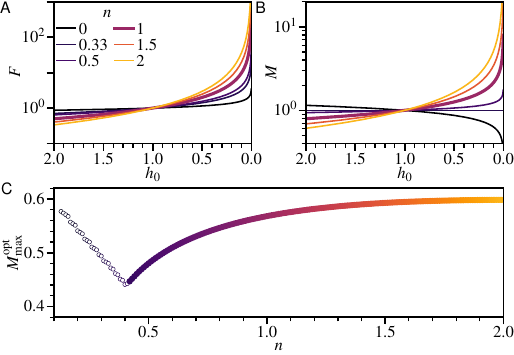}
    \caption{Predicted impact response for a power-law fluid. \textit{(A)}~Constant velocity impact force, $F(h_0)$, at different index, $n$, dark (purple) thinning to light (yellow) shear-thickening (see legend). Normalised by setting additional parameters to unity. \textit{(B)}~Corresponding bending moment, $M(h_0)$. \textit{(C)}~Peak $M(t)$ for decelerating impact vs power-law index, $M^{\rm opt}_{\max}(n)$, at optimal consistency, $K$, following \partFigref{fig:newtonian}{D}. Symbols: light (thickening) to dark (thinning); open, impact to $h_0(t) < 10^{-4}$.}
    \label{fig:powerlaw}
\end{figure}

Consider first a constant-speed impact. We plot in \partFigref{fig:powerlaw}{A--B} the $h_0(t)$ dependence implied by \eqref{eq:wFsolved}:
\begin{equation}
    F \propto  h_0^{-\frac{5n+1}{n+5}}~\textrm{and}~M = F\weff \propto h_0^{-\frac{3n-1}{n+5}}. \label{eq:heightNN}
\end{equation}
The force and bending moment in a shear-thickening fluid laminate ($n = 1.5$, 2) diverge more sharply as the gap narrows than the Newtonian case ($n=1$). However, a shear thinning fluid ($n = 0.5, 0.33, 0$) leads to a weaker force divergence. For $n= 0.5$ the bending moment also diverges more weakly than the Newtonian case. Interestingly, decreasing $n$ further brings a constant $M$ ($n = 0.33)$ and then a decreasing $M$ ($n = 0$). These results suggests that for laminate protection a shear-thinning, \emph{not thickening}, fluid is needed.

We next confirm and generalise our analysis with numerical solutions of the dynamical equation for $h_0(t)$:
\begin{equation}
    \frac{\rmd^2 h_0}{\rmd t^2} = - \frac{\sqrt{12Kv}}{h_0}\left(\frac{(6v)^5}{2Kh_0^8} \right)^\frac{n-1}{2(n+5)}
    \label{eq:impact}
\end{equation}
where the second term modifies the Newtonian equation, \eqref{eq:NewtonianImpact}, and $B$, $L$ and $m$ have been set to unity. 

For any value of $n\gtrsim 0.4$, we find an optimal $K$ for which the maximum bending moment is minimised (comparable to \partFigref{fig:newtonian}{D}, but with $\eta \to K$). Increasing $n$ from the Newtonian value of unity, this optimal value $\Mopt$ increases, \partFigref{fig:newtonian_impact}{C}, \ie, a shear-thickening fluid decreases protection. In contrast, decreasing $n$ below unity, \ie, changing to progressively more shear-thinning fluids, lowers $\Mopt$, thus offering increasing impact protection, consistent with our constant-$v$ analysis.  

For $n<0.4$, we find that decreasing $K$ below its optimal value brings laminate failure, as $h_0 \to 0$.  So, we predict that optimal impact protection is offered by a shear-thinning fluid with $n = 0.4$, somewhat higher than the $\frac{1}{3}$ from the constant-$v$ analysis, but is insensitive to pre-factors in our scaling analysis. Physically, a shear-thinning fluid is optimal as it is harder to push out of large gaps (low $\dot\gamma$, higher $\eta_{\rm eff}$, larger $F$) than for narrow gaps (high $\dot\gamma$, lower $\eta_{\rm eff}$, smaller $F$), which smooths $F(t)$ and hence $M(t)$.

\subsection*{Constant velocity experiments}%
\begin{figure}[tb]
    \centering
    \includegraphics[width=\columnwidth]{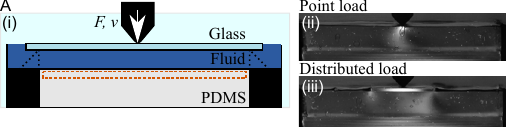}
    \includegraphics[width=\columnwidth]{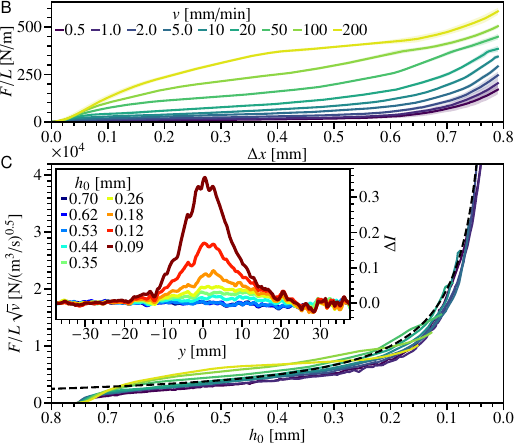}
    \caption{Experimental controlled-velocity impact into a Newtonian fluid laminate. \textit{(A)}~Testing apparatus. \textit{(i)}~Diagram from top to bottom: \SI{0.3}{\milli\metre} glass; \SIrange{0.70}{0.76}{\milli\metre} fluid layer; base with \SI{10}{\milli\metre} PDMS, region analysed for pressure measurement, dashed (orange) outline. Fluid flow out of plane prevented by rigid glass panes (light shading); laterally serrated anvils allow fluid flow during loading, SI Appendix Fig.~S2. \textit{(ii)}~Image, $I(x,y)$, of static point loading, $F/L\approx \SI{200}{\newton\per\metre}$, directly on PDMS in dark-field circular polariscope. Note, background subtraction has not been performed and the air pocket created is unique to the localised load directly on the PDMS. \textit{(iii)}~Distributed static load, across \SI{20}{\milli\metre} rigid glass slide. \textit{(B)}~Force-displacement response with varying speed, $v$, for \SI{0.3}{\milli\metre} thick glass with \SI{0.76}{\milli\metre} initial gap. Lines: dark (purple) to light (yellow), slow to fast controlled $v$ (see inset legend), three test average, standard deviation shown by shading. \textit{(C)}~Velocity-normalised force, $F/L\sqrt{v}$, as a function of corrected gap, $h_0$, line shading as in (B). Dashed black line: model prediction, $F/L = 12(\eta v B/L)^{1/2}/h_0$. Inset: polariscope proxy pressure measurement. Vertically averaged intensity change, $\Delta I(y)$, across quasi-2D geometry at decreasing $h_0$ [blue to dark (red), see inset legend]. Impact velocity, $v=$\,\SI{20}{\milli\per\minute} and $h_i =$\,\SI{0.7}{\milli\metre}. }
    \label{fig:newtonian_impact}
\end{figure}
We verify our analysis in an experimental realisation of our quasi-2D set up from \partFigref{fig:fsi}{A}, using a universal testing machine to drive a wedge downwards at a laminate consisting of a fluid sandwiched between a \SI{0.3}{\milli\meter}-thick flexible glass plate and a \SI{10}{\milli\meter}-thick polydimethylsiloxane (PDMS) base, \partFigref{fig:newtonian_impact}{A}, at low enough constant velocity, $v$, to allow us to follow the force on the wedge, $F$, as a function of time, or, equivalently, (downward) displacement, $\Delta x$. The gap height is $h_0 = \hin - \Delta x + F/k$, where $\hin$ is the initial gap height, and $k$ is the (separately measured) stiffness of the system. We measured $F(\Delta x)$ at different imposed $v$, and monitored the pressure on the PDMS via photoelastic imaging. Experimental details are in Materials and Methods. 

\subsubsection*{Newtonian fluids}

We begin with a Newtonian fluid laminate with $\hin = \SI{0.7}{\milli\meter}$, using glycerol as the `sandwich filling', increasing $v$ from \SI{0.5}{\milli\metre\per\minute}, \partFigref{fig:newtonian_impact}{B}~[dark (purple) lines], to \SI{200}{\mmpm} [light (yellow) lines]. At low $v$, the fluid can almost freely drain and $F$ is low, only increasing as $\Delta x \to \SI{0.8}{\milli\metre}$ and $h_0 \to 0$. With increasing $v$, $F(\Delta v)$ takes on a sigmoidal shape. Converting $\Delta x$ to $h_0$ and normalising by $\sqrt{v}$ collapses the data to within a factor of 1.5 over a 400-fold variation in $v$, \partFigref{fig:newtonian_impact}{C}. confirming the $\sqrt{v}$ scaling of \eqref{eq:FWnewt}. Indeed, $F/L = 12(\eta v B/Lh_0^2)^{1/2}$ offers a credible account of the collapsed data (dashed line). That this is within an order-unity numerical factor ($\sqrt{12} \approx 3.5$) of \eqref{eq:FWnewt} validates the physics embodied in our scaling analysis: an effective squeeze flow that shrinks in extent as the viscous forces more strongly bend the flexible upper layer. 

To illustrate this physics, we turn to photo-elastic measurements, where light intensity is a proxy for the pressure, so that we can visually distinguish between a point and a distributed load, \partFigref{fig:newtonian_impact}{A}~(ii) and (iii) respectively. At $v=\SI{20}{\milli\metre\per\minute}$, a bright region, evidencing high pressure, emerges at $h_0 \lesssim \SI{0.35}{\milli\metre}$ \partFigref{fig:newtonian_impact}{C}, and grows in intensity as $h_0$ decreases further. The half width of a Gaussian fitted to the measured intensity pattern decreases only weakly, from 9.9(2) to \SI{6.19(3)}{\milli\metre} as $h_0$ decreases from 0.53 to \SI{0.09}{\milli\metre}. The observation of a localised high pressure region is consistent with assumption of squeeze flow in a confined region of some effective width $\weff$. The weak dependence of $\weff$ on $h_0$ is also consistent with \eqref{eq:FWnewt}, from which we predict $\weff \simeq (\B h_0^4/12L\eta v)^{1/6} = \SI{9}{\milli\metre}$ at $h_0 = \SI{0.35}{\milli\metre}$ down to $\weff \simeq \SI{4}{\milli\metre}$ at $h_0 = \SI{0.09}{\milli\metre}$, comparable to the observed widths and trends of the high-pressure region. Finally, these results are  consistent with our assumptions of lubrication flow ($\weff \gg h_0$) and neglecting boundaries ($\weff \ll W = \SI{75}{\milli\metre}$). Thus, the complex feedback between fluid flow and plate deformation can indeed be captured in an `effective flat plate' treatment.

\subsubsection*{Non-Newtonian fluids}

\begin{figure}[tb]
    \centering
    \includegraphics[width=\columnwidth]{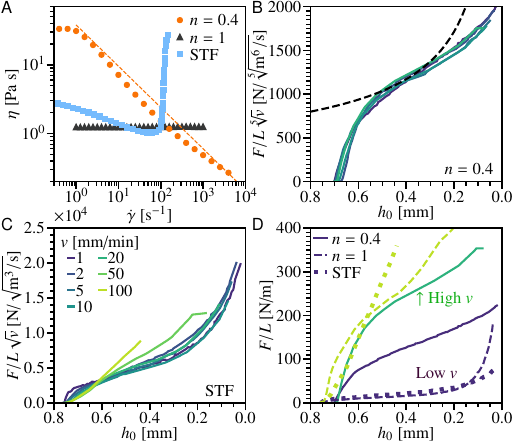}
    \caption{Impact into non-Newtonian fluid based laminates. \textit{(A)}~Fluid rheology, viscosity with shear rate, $\eta(\dot\gamma)$. Symbols: light (blue) squares, shear-thickening suspension of 20\,wt\% fumed silica in PEG200 measured with fixed stress; (orange) circles, 7\,wt\% suspension of hydrophobic fumed silica in PEG200 measured at fixed rate; and dark (grey) triangles, glycerol. Dashed line, representative power-law fit for shear-thinning region, $\eta = K\dot\gamma^{n-1}$, for $n=0.4$, $K=38$~Pa\,s$^{0.4}$ and $\dot\gamma=$~1 to $10^4$\,s$^{-1}$. \textit{(B)}~Shear-thinning fluid. Force, $F(h_0)/L$, normalised by speed, $\sqrt[5]{v}$, for $\hin = \SI{0.70}{\milli\metre}$. Lines: solid, dark (purple) to light (green), $v = $~\SIrange{1}{20}{\mmpm}, see legend in part B; dashed, model prediction [\eqref{eq:wFsolved}, $n=0.4$, 2.4 pre-factor]. \textit{(C)}~Shear-thickening fluid. Force normalised for Newtonian fluid by $\sqrt{v}$. Lines, increasing $v$, see inset legend. \textit{(D)}~Comparison of force response for different fluid rheologies. Dark lines, low speed, $v=\SI{1}{\mmpm}$: thin solid, shear thinning; dashed, Newtonian; and thick dotted, STF. Light lines, high $v$, $v=20$, 50 and \SI{100}{\mmpm} respectively.}
    \label{fig:nn_impact}
\end{figure}

We next tested a laminate filled with an $n = 0.4$ shear-thinning suspension, \partFigref{fig:nn_impact}{A} (filled circles); this and the shear-thickening suspension (see below) can be treated as continua, as the particle size is much smaller than the minimum gap (SI Appendix). Now, \eqref{eq:wFsolved} predicts $F \propto v^{0.22}$, consistent with the observed collapse of $F(h_0)$ data taken at different speeds when we plot $F(h_0)/\sqrt[5]{v}$, \partFigref{fig:nn_impact}{B}. The prediction of $F \propto h_0^{-0.55}$ [\eqref{eq:heightNN}] does not capture the transient, early-stage response, but shows moderate agreement at intermediate $h_0$, \partFigref{fig:nn_impact}{B}~(dashed), with a prefactor of 2.4 consistent with a scaling analysis. The observed divergence in $F$ as $h_0 \to 0$ is weaker than for $n=1$, matching the predicted trend. However, it is also weaker than predicted  for $n=0.4$. Better agreement between theory and experiment here may require more careful modelling of shear-thinning fluids under squeeze flow conditions~\cite{meeten2005flow}.

If instead a shear-thickening fluid, \partFigref{fig:nn_impact}{A} (filled squares), is used, we observe a markedly different behaviour. Varying $v$ from \SIrange{1}{20}{\mmpm}, \partFigref{fig:nn_impact}{C}~[dark (purple) to light (green)], we find that $F(h_0)$ is Newtonian-like, with $F/\sqrt{v}$ collapsing the data (\cf\ \partFigref{fig:newtonian_impact}{C}). This is consistent with the almost-constant viscosity of this fluid at low shear rates: $\eta$ decreases from $3$ to $\SI{1}{\pascal\second}$ as $\dot\gamma$ increases from $10^{-1}$ to $10^2\,\si{\second^{-1}}$. A different behaviour is seen when $v \geq \SI{50}{\mmpm}$, \partFigref{fig:nn_impact}{C}~(light lines): $F/\sqrt{v}$ no longer collapses the data, and  the $h_0$ dependence becomes stronger, although the small-$h_0$ limit could not be accessed in these high $v$ experiments due to load cell limits. The shear rate at the onset of this change can be estimated by using \eqref{eq:FWnewt} for $\weff$ with $\eta = \SI{1}{\pascal\second}$, so that $\dot\gamma = 6v\weff/h_0^2 \sim $~\SI{160}{\per\second} at $v = $~\SI{50}{\milli\metre\per\minute} and $h_0=$~\SI{0.6}{\milli\metre}. This is consistent with the shear rate at which we observe shear thickening in our fluid, \partFigref{fig:nn_impact}{A} (filled squares), once again supporting the validity of our analysis in terms of an effective flat plate of width $\weff$, and an effective viscosity set by the edge shear rate, $\etaf = \eta(\rim)$.

\subsection*{Energy scaling}

The speeds at which we have performed our experiments to validate our scaling analysis are far too low for realistic impact protection at $v \gtrsim \SI{1}{\metre\per\second}$. Nevertheless, our analysis, now substantially validated by experiments, allows some predictions for higher speeds via energy scaling. 

The kinetic energy scales as $v^2$, but $F$ (and energy absorbed) scales as $v^{0.2}$  for the optimal-protection shear thinning fluid with $n=0.4$, \eqref{eq:wFsolved}. For a  laminate with given ($\hin$, $B$), the consistency $K$ required for energy absorption increases with $v$. In a constant-$v$ approximation,
\begin{equation}
    \int_0^{\hin} \! \rmd h_0\, \frac{F}{L} \sim  \frac{n+5}{4(1-n)} \hin ^\frac{4(1-n)}{n+5}\sqrt{\frac{vKB}{L}} \left( \frac{v^5 B}{KL} \right)^\frac{n-1}{2(n+5)}\!\!.
\end{equation} 
So, for energy $\sim$~\SI{0.25}{\joule} (\eg, $m = \SI{50}{\gram}$ for $L = \SI{25}{\milli\metre}$ and $v=\SI{3}{\metre\per\second}$), our model laminate ($B/L = \SI{0.18}{\newton\metre}$, $\hin = \SI{1}{\milli\metre}$) requires $K \sim 10^4$\,Pa\,s$^{0.4}$. For this fluid, even a low $\dot\gamma \sim \SI{1}{\per\second}$ would generate stresses $\sim10^4$\,Pa.

Under such conditions, our fumed silica suspensions may become brittle~\cite{di2012rheological}, rendering manufacturing challenging,  and post-impact `self healing' may not be possible. A fluid with more complex rheology, \eg, one that thins only at the high $\dot\gamma$ of impact, may be more suitable. This reduces stresses at slow deformation, facilitating manufacturing, self-healing, and, perhaps, even enabling fully flexible laminates. Such rheology could be achieved using suspensions that thin after thickening, due to asperity compression~\cite{lobry2019shear} or a brush-like coating~\cite{nguyenle2022solvents}, or a polymer solution with a low-shear plateau~\cite{ryder2006shear}.

\section*{Conclusions}

Inspired by the use of shear-thickening fluids in body armours, we have established a general scaling framework for analysing the impact response of solid-fluid laminates, which captures interactions through an effective rigid plate squeeze flow with width $\weff$, which scales only weakly with all parameters, \eqref{eq:FWnewt}. Insight can, therefore, be gained by thinking in terms of a simple rigid plate squeeze flow. Strikingly, we conclude that, not thickening, but shear thinning with $\eta \propto \dot\gamma^{-0.6}$ optimises protection, \partFigref{fig:nn_impact}{D}. This arises from reducing the $F(h_0)$ divergence, with a low $\eta_{\rm eff}$ at small $h_0$ (high $\dot\gamma$), while still absorbing the impact energy with a high $\eta_{\rm eff}$ at large $h_0$ (smaller $\dot\gamma$). These scaling predictions were substantially verified in controlled-velocity impact tests where we measured $F(h_0)$ and imaged the pressure distribution using photoelasticity. Together, these results establish the effective rigid plate squeeze flow approximation as a useful tool for analysing fluid-solid interactions in composites incorporating non-Newtonian fluids. 

Further work including flow perpendicular to $x$ and $y$~\cite{laun1999analytical} or curvature~\cite{meeten2005flow}, as well as normal stress differences~\cite{royer2016rheological}, strain-dependence~\cite{richards2019competing} and extensional viscosities~\cite{cheal2018rheology}, could allow predictive design of optimised fluids for realistic impact velocities. More generally, our scaling approach may also apply to non-Newtonian fluid-solid interaction problems arising from rubbing skin ointments~\cite{cyriac2022tactile} or eating chocolate~\cite{rodrigues2021frictional}.

\section*{Materials and Methods}

\small{Non-Newtonian fluids were prepared from fumed silica in poly-ethylene glycol (PEG 200, Sigma Aldrich), with a shear-thinning suspension from 7\,wt\% hydrophobic hexamethyldisilazane-modified Aerosil\textregistered~R812S and a shear-thickening suspension from 20\,wt\% hydrophilic HDK\textregistered~N20. Particles are $\sim$\,\SI{100}{\nano\metre} radius (Fig.~S3) fractal-like aggregates~\cite{ibaseta2010fractal} of $\approx$\,\SI{3}{\nano\metre} primary particles. Powders were dispersed via vortex mixing, then repeated stirring and centrifugation to break agglomerates~\cite{kamaly2017dispersion}, similar to conching~\cite{blanco2019conching}.

Rotational rheometry (NETZSCH Kinexus Ultra+) was performed at $T = \SI{20}{\celsius}$. For the shear-thickening fluid, controlled-stress measurements were made with roughened parallel plates (radius, $R=\SI{10}{\milli\metre}$ and gap, $H = \SI{200}{\micro\metre}$); we report the rim shear rate, $\dot\gamma = \Omega R/H$, from the measured rotation rate and the viscosity based on the apparent stress, $\sigma = 2\mathcal{T}/\pi R^3$, from the applied torque, \partFigref{fig:nn_impact}{A}~(blue squares). Stress was applied from \SI{1}{\pascal} logarithmically at 10 pts/decade with \SI{10}{\second} equilibration and \SI{10}{\second} measurement at each point up to the fracture stress (\SIrange{3}{10}{\kilo\pascal}), ensuring reversibility in separate tests. For the shear-thinning fluid, rate-controlled measurements were made in a smooth cone-plate geometry (angle, $\alpha = \SI{1}{\degree}$ angle; $R = \SI{20}{\milli\metre}$) with $\dot\gamma = \Omega/\sin(\alpha)$ and $\sigma = 3\mathcal{T}/2\pi R^3$, \partFigref{fig:nn_impact}{A}~(orange circles). Shear rates were applied at 5~pts/decade from $\dot\gamma = \SI{0.01}{\per\second}$ to inertial ejection, $\dot\gamma = \SI{4000}{\per\second}$. For glycerol (99\,wt\%, Fisher Scientific), measurements were made at 10~pts/decade from \SIrange{1}{1000}{\per\second}, \SI{5}{\second} equilibration and \SI{10}{\second} measurement.

Viscosities are shown relative to Newtonian glycerol (\partFigref{fig:nn_impact}{A} grey triangles, $\eta = \SI{1.24}{\pascal\second}$). Hydrophilic silica initially weakly shear thins, before reaching a critical rate, $\dot\gamma_c \simeq \SI{100}{\per\second}$, where further stress does not increase the rate (discontinuous shear thickening~\cite{barnes1989shear}). This is consistent with previous results~\cite{van2021role}, with the onset of thickening occurring when the stabilising force, attributed to the absorption of PEG onto the silica surface, is overcome and the particles enter frictional contact~\cite{lin2015hydrodynamic}. Compared to monodisperse spheres, DST occurs at a low volume fraction, $\approx 11\%$, which may be attributed to the fractal-like nature of the particles with additional rolling constraints~\cite{guy2018constraint,singh2020shear}.

Hydrophobic silane surface modification creates a strongly shear-thinning material~\cite{raghavan1998composite}, \partFigref{fig:nn_impact}{A}~(orange circles), similar to removing adsorbed surfactants~\cite{richards2021turning}. At low $\dot\gamma$ slip is observed~\cite{walls2003yield}, above this shear thinning with $n \approx 0.4$ (dashed line, $K = 38$~Pa\,s$^{0.4}$) occurs up to sample fracture. Around $\dot\gamma=\SI{100}{\per\second}$, $\eta$ for all fluids are comparable, at the range of $\dot\gamma$ for low-velocity impact testing. The three fluids, with comparable absolute $\eta$ but different $\dot\gamma$ dependence, allow isolation of the role of fluid rheology. 

Our quasi-2D controlled-velocity impact apparatus is based on a universal testing machine (Lloyd Instruments LS5, AMETEK). The force-displacement response (20 or \SI{100}{\newton} load cell, \SI{1}{\kilo\hertz} sampling) is measured with $v = \SIrange{0.5}{200}{\milli\metre\per\minute}$. Combined with a dark-field circular polariscope (FL200, G.U.N.T.\ Ger\"atebau GmbH) and a photo-elastic base, qualitative pressure measurements can be made.

Our top flexible plate, \partFigref{fig:newtonian_impact}{A}, was $\SI{25}{\milli\meter}\times \SI{75}{\milli\meter} \times \SI{0.3}{\milli\meter}$ glass. The base was a \SI{10}{\milli\metre}-thick piece of cut silicone elastomer [Sylgard 184, Dow Chemical Company, 5:1 cross-linker ratio, degassed and cured at \SI{25}{\celsius} for \SI{48}{\hour}, $E=$~\SI{1.5}{\mega\pascal}~\cite{johnston2014mechanical}]. The constraining panels were sealed with silicone oil (10,000\,cSt, Sigma Aldrich). For non-Newtonian fluid force-displacement tests, glass was on top of the base (compliance, $k = \SI{80}{\newton\per\metre}$, $\hin = \SI{0.76}{\milli\metre}$); otherwise $k = \SI{50}{\newton\per\milli\metre}$, $\hin = \SI{0.7}{\milli\metre}$.

For force-displacement measurements, the initial gap, $\hin$, and zero displacement, $\Delta x =0$, were set with no fluid. After loading the fluid, the laminate was allowed to come to equilibrium, $F=0$ and $\Delta x = 0$. The impactor was then moved down \SI{0.8}{\milli\metre} at a fixed speed, $v$, recording $F(t)$ and $\Delta x (t)$ from which $F(\Delta x)$ was reconstructed. The gap, $h_0 = \hin - \Delta x + F/k$.

To infer the fluid pressure, we used a polariscope to probe stress in the base, giving finer spatial resolution than transducer arrays~\cite{grandes2021rheological,gauthier2021new}. Stress-induced intensity patterns in the PDMS, $I(x,y,t)$, were recorded using a camera (Nikon Z6, $3840\times2160$ \SI{30}{\hertz}, 8-bit grey-scale). Instead of precisely quantifying the stress~\cite{dally1991experimental}, we sought to establish the extent of any high-pressure region. A narrow region at the top of the base layer is isolated in recording, $700\times10$~px$^2$, \partFigref{fig:newtonian_impact}{a}~(red outline). The change in intensity from the quiescent state at the start of recorded movies, $\Delta I(x,y,t)$, is averaged vertically, $\Delta I(y,t)$, and smoothed on short length scales using a Savitzky-Golay filter. The intensity is normalised to saturation (ISO 1200 and shutter speed 1/125).}

\section*{Acknowledgements}

We thank Alexander Morozov and Patrick Warren for fruitful discussions. Edinburgh work was partly supported by the UK Engineering and Physical Sciences Research Council Impact Acceleration Account (Grant No.\ EP/R511687/1). {Author contributions:} J.A.R., D.J.M.H., M.E.D., and W.C.K.P.\ designed the research; J.A.R.\ and R.E.O'N.\ developed the experimental methods; J.A.R.\ performed experiments and developed the model; D.J.M.H.\ and W.C.K.P.\ supervised the research; all authors discussed and interpreted the results; J.A.R.\ and W.C.K.P.\ wrote the paper; and all authors commented on the manuscript. 

\section*{References}

%

\pagebreak
\widetext
\begin{center}
\textsf{\textbf{\Large Supplemental Information: Optimising non-Newtonian fluids for impact protection of laminates}}
\end{center}
\setcounter{equation}{0}
\setcounter{figure}{0}
\setcounter{table}{0}
\setcounter{page}{1}
\makeatletter
\renewcommand{\theequation}{S\arabic{equation}}
\renewcommand{\thefigure}{S\arabic{figure}}
\renewcommand{\thesection}{S\Roman{section}}
\renewcommand{\bibnumfmt}[1]{$^{\rm S#1}$}
\renewcommand{\thepage}{S\arabic{page}} 

\twocolumngrid

\section{Effect of curvature}

\begin{figure}[h]
\centering
\includegraphics{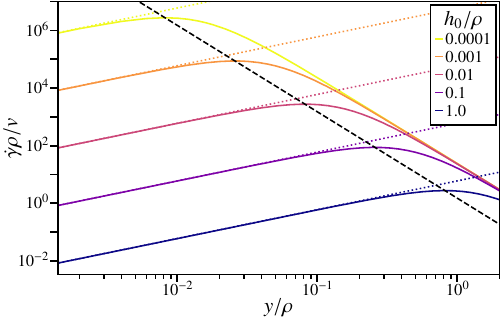}
    \caption{Impact of curvature on wall shear rate. Shear rate normalised by velocity and radius of curvature, $\dot\gamma v/\rho$, as a function of normalised distance from point of impact, $y/\rho$. Lines: solid, shear rate for narrowing gap (dark to light, see inset legend for $h_0/\rho $ values); dotted, shear rate for flat plate, $h = h_0$; and, dashed, $y/\rho = \sqrt{2h_0/3\rho}$, defining maximum shear rate and scale for transition from high pressure region at $\Delta h = y^2/2\rho = h_0/3$.}
    \label{fig:curve}
\end{figure}

In the main text, we define an effective width $\weff$ to denote the region in which there are significant pressure gradients. Since $\partial_y P \sim h_0^{-3}$, Eq.~(2), this region is delimited by the criterion of $\Delta h \lesssim h_0$. Over the same region, we have claimed that it is reasonable to approximate the top plate as flat, because it is only weakly curved. Here, we demonstrate the truth of this claim via calculating the shear rate. The same calculation will be relevant when we consider non-Newtonian fluids. 

For a rigid curved plate, the gap can be approximated by $h(y)=h_0+y^2/2\rho$, for a radius of curvature, $\rho$. With $Q = vy$ (unchanged from a flat plate) by conservation of volume, the wall shear rate is 
\begin{equation}
    \label{eq:curve}
    \dot\gamma = \frac{6Q}{h^2} = \frac{6vy}{\left( h_0 + \frac{y^2}{2\rho}\right)^2}\,.
\end{equation}
Normalising lengths by $\rho$ and plotting the shear rate with $y$, Fig.~\ref{fig:curve}, demonstrates that the shear rate for small $y$ is equivalent to a flat plate, \textit{cf}.\ solid and dotted lines. Here $h\approx h_0$ and $\dot\gamma \propto vy/h_0^2$. At large $y$, the growth in $h$ dominates and the shear rate drops, $\dot\gamma \propto \rho^2v/y^3$. For narrower gaps, or more curved surfaces $h_0/\rho\ll 1$, the flat-plate--like region where shear rate is $\propto y$ becomes smaller (dark to light lines). The transition defined by $\max(\dot\gamma)$, occurs at $y = \sqrt{2h_0\rho/3}$, Fig.~\ref{fig:curve}~(dotted line). At this point the gap has increased by $h_0/3$. Within a numerical constant of order unity, this is the same criterion as that we have used to define $\weff$ ($\Delta h \sim h_0$) based on Eq.~(2) alone and used in our scaling analysis. The shear rate in the curved case is $9/16\approx 0.56$ times that for a plate, while the pressure gradient is $\approx 0.42$ that for a flat plate.

\begin{figure}[t]
    \centering
    \includegraphics[width=0.5\textwidth]{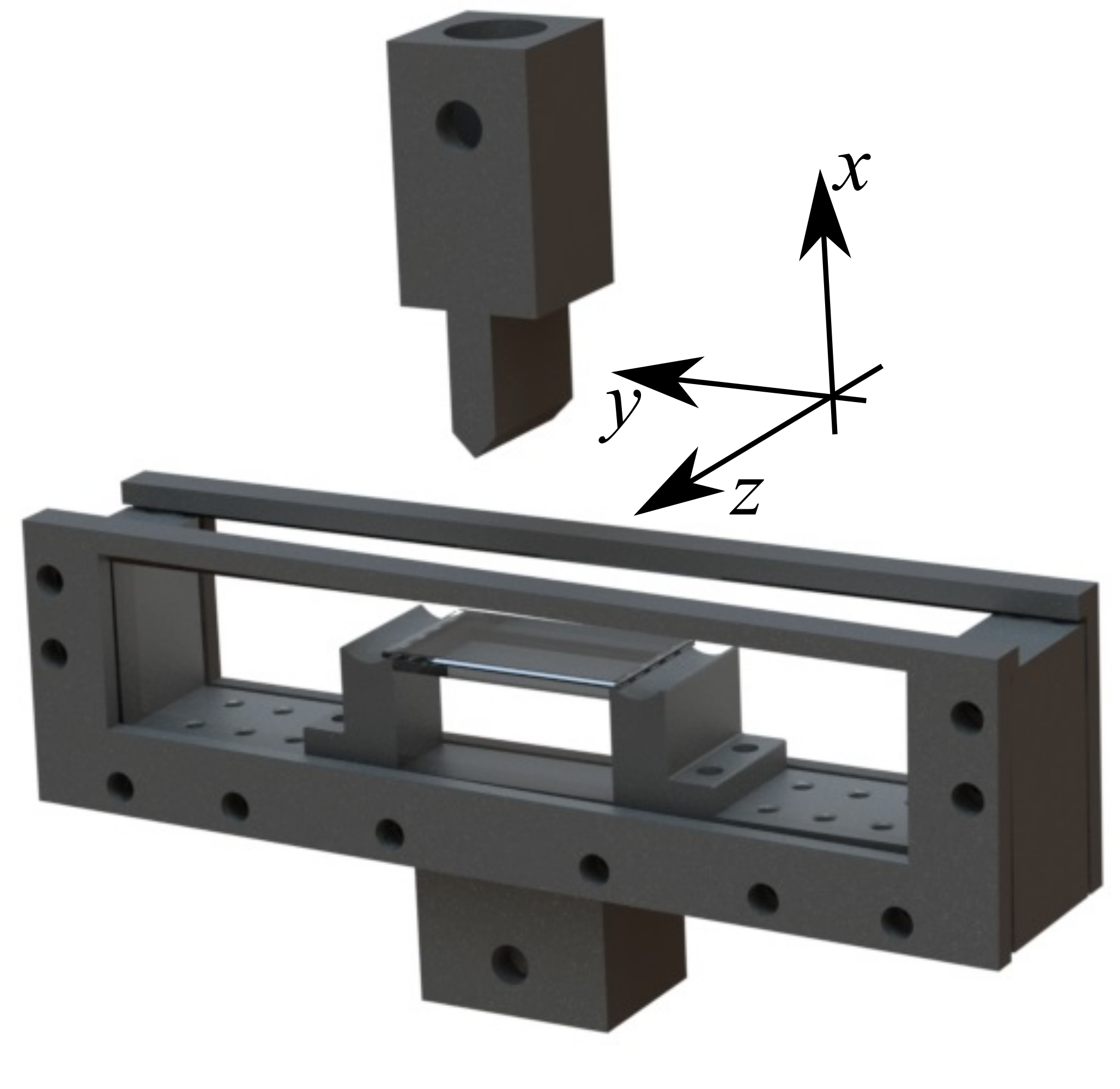}
    \caption{3D rendering of quasi-2D laminate geometry and impactor showing movable support points for flexible glass layer and confining panels to prevent flow along $z$ direction.}
    \label{fig:3d}
\end{figure}

\section{Derivation of closure relation for non-Newtonian fluids\label{sec:deriv}}

To describe the non-Newtonian response of the fluid we substitute $\weff$, Eq.~(5), into $\rim \simeq 6v\weff/h_0^2$ to obtain
\begin{equation}
    \rim \simeq 6\left( \frac{Bv^5}{12\eta h_0^8 L} \right)^{1/6}.
\end{equation}
Substituting $\eta = K\dot\gamma^{n-1}$ and solving for $\rim$, we then find
\begin{equation}
    \rim \simeq 6^{\frac{6}{n+5}}\left( \frac{Bv^5}{12Kh_0^8 L} \right)^{\frac{1}{n+5}},
\end{equation}
so that the effective Newtonian viscosity is
\begin{equation}
    \eta_{\rm eff} = K \rim^{n-1} \simeq 6^{\frac{6(n-1)}{n+5}}K\left( \frac{Bv^5}{12Kh_0^8 L} \right)^{\frac{n-1}{n+5}}
\end{equation}
As in our reduced Newtonian FSI solution, $\eta_{\rm eff}$ sets the effective width of the squeeze flow region. From Eq.~(5), we then find
\begin{equation}
    \weff \simeq \left( \frac{Bh_0^4}{12\eta_{\rm eff} v L} \right)^{\frac{1}{6}} \simeq 6^{\frac{1-n}{n+5}}\left( \frac{Bh^4}{12K v L} \right)^{\frac{1}{6}}\left( \frac{Bv^5}{12Kh_0^8 L} \right)^{\frac{1-n}{6(n+5)}}.
\end{equation}
This sets the force that stops the impact and bends the glass. Again, from Eq.~(5), we find
\begin{equation}
    \frac{F}{L} \simeq \frac{12\eta_{\rm eff}v\weff^3}{h_0^3} \simeq \frac{12\eta_{\rm eff} v}{h_0^3} \left(\frac{\B h_0^{4}}{12\eta_{\rm eff}vL}\right)^{\frac{1}{2}} = \frac{\sqrt{12}}{h_0} \left( \frac{vB}{L}\right)^{\frac{1}{2}}\eta_{\rm eff}^{\frac{1}{2}},
\end{equation}
from which,
\begin{equation}
    \frac{F}{L} \simeq \frac{\sqrt{12} \times 6^{\frac{3(n-1)}{n+5}}}{h_0} \left( \frac{KvB}{L}\right)^{\frac{1}{2}} \left( \frac{Bv^5}{12Kh_0^8 L} \right)^{\frac{n-1}{2(n+5)}}.
\end{equation}
Again, we recover the Newtonian case if $n= 1$, $K = \eta$.

\section{Particle sizing\label{sec:sizing}}

Throughout, we have treated our experimental non-Newtonian fluids as continua. This requires that the minimum gap size we resolve during impact, $\approx \SI{50}{\micro\meter}$ (Fig.~4C), always remains significantly larger than the size of the particles and agglomerates in our suspensions. We have therefore measured the particle size distribution (PSD) of our experimental suspensions to ensure proper dispersal before use. 

\begin{figure}[htb]
    \centering
    \includegraphics{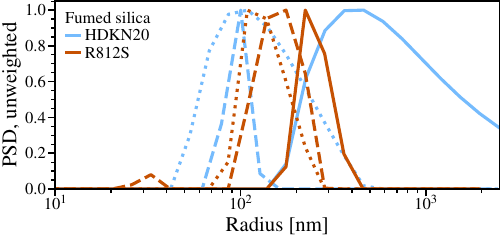}
    \caption{Particle size distribution (PSD) from dynamic light scattering for hydrophilic (HDK N20, blue) and hydrophobic (R812S, red) fumed silica for various dispersal methods (lines: solid, dilute shear; dashed, sonicated; and dotted, concentrated shear and dilution). Particle size distribution normalised to peak value, presented as unweighted results from regularised exponential analysis.}
    \label{fig:sizing}
\end{figure}

Dilute suspensions ($\varphi = 0.05$\,wt\%) were prepared in a compatible solvent, distilled water (N20) or ethanol (R812S). The PSD was then determined using dynamic light scattering (DLS, ALV LSE-5004 at \SI{632.8}{\nano\metre} and \SI{45}{\degree} scattering angle) using a regularised fit of the decorrelation function, Fig.~\ref{fig:sizing} [N20, light (blue) lines and R812S, dark (orange) lines]. This is presented unweighted as fumed silica is a fractal-like particle and the standard radius dependence does not apply~\cite{ibaseta2010fractal}. Fumed silica was dispersed via three methods. Firstly via simple vortex mixing that leaves agglomerates intact~\cite{kamaly2017dispersion}, solid lines. Secondly, when dispersed with further high energy input sonication (Sonics vibra-cell, \SI{500}{\watt} power with a tapered probe at 20\% power with \SI{10}{\second} pulses for \SI{2}{\minute}), dashed lines. Finally, samples as prepared for laminate testing were diluted and vortex mixed to $\varphi = 0.05$\,wt\%, dotted lines.

For particles mixed using dilute shear, the PSD span is broad, being spread from \SI{150}{\nano\metre} upwards, peaking at \SI{220}{\nano\metre} for R812S and \SI{410}{\nano\metre} for N20, then with a tail of large radius particles up to \SI{0.5}{\micro\metre} or \SI{2.5}{\micro\metre} respectively, Fig.~\ref{fig:sizing}~(solid lines). When particles are dispersed via high-energy probe ultrasound, a narrower PSD is found with peaks at \SI{100}{\nano\metre} and \SI{150}{\nano\metre}~(dashed lines) associated with the diameter of the permanently fused aggregates of primary particles. Comparable peaks are found with high volume fraction shear followed by dilution, dotted lines. This indicates that the mixing method used for suspension preparation breaks up agglomerates.




\begin{thebibliography}{47}%
\makeatletter
\providecommand \@ifxundefined [1]{%
 \@ifx{#1\undefined}
}%
\providecommand \@ifnum [1]{%
 \ifnum #1\expandafter \@firstoftwo
 \else \expandafter \@secondoftwo
 \fi
}%
\providecommand \@ifx [1]{%
 \ifx #1\expandafter \@firstoftwo
 \else \expandafter \@secondoftwo
 \fi
}%
\providecommand \natexlab [1]{#1}%
\providecommand \enquote  [1]{``#1''}%
\providecommand \bibnamefont  [1]{#1}%
\providecommand \bibfnamefont [1]{#1}%
\providecommand \citenamefont [1]{#1}%
\providecommand \href@noop [0]{\@secondoftwo}%
\providecommand \href [0]{\begingroup \@sanitize@url \@href}%
\providecommand \@href[1]{\@@startlink{#1}\@@href}%
\providecommand \@@href[1]{\endgroup#1\@@endlink}%
\providecommand \@sanitize@url [0]{\catcode `\\12\catcode `\$12\catcode
  `\&12\catcode `\#12\catcode `\^12\catcode `\_12\catcode `\%12\relax}%
\providecommand \@@startlink[1]{}%
\providecommand \@@endlink[0]{}%
\providecommand \url  [0]{\begingroup\@sanitize@url \@url }%
\providecommand \@url [1]{\endgroup\@href {#1}{\urlprefix }}%
\providecommand \urlprefix  [0]{URL }%
\providecommand \Eprint [0]{\href }%
\providecommand \doibase [0]{https://doi.org/}%
\providecommand \selectlanguage [0]{\@gobble}%
\providecommand \bibinfo  [0]{\@secondoftwo}%
\providecommand \bibfield  [0]{\@secondoftwo}%
\providecommand \translation [1]{[#1]}%
\providecommand \BibitemOpen [0]{}%
\providecommand \bibitemStop [0]{}%
\providecommand \bibitemNoStop [0]{.\EOS\space}%
\providecommand \EOS [0]{\spacefactor3000\relax}%
\providecommand \BibitemShut  [1]{\csname bibitem#1\endcsname}%
\let\auto@bib@innerbib\@empty
\bibitem [{\citenamefont {Lee}, \citenamefont {Wetzel},\ and\ \citenamefont
  {Wagner}(2003)}]{lee2003ballistic}%
  \BibitemOpen
  \bibfield  {author} {\bibinfo {author} {\bibfnamefont {Y.~S.}\ \bibnamefont
  {Lee}}, \bibinfo {author} {\bibfnamefont {E.~D.}\ \bibnamefont {Wetzel}},\
  and\ \bibinfo {author} {\bibfnamefont {N.~J.}\ \bibnamefont {Wagner}},\
  }\bibfield  {title} {\enquote {\bibinfo {title} {The ballistic impact
  characteristics of Kevlar{\textregistered} woven fabrics impregnated with a
  colloidal shear thickening fluid},}\ }\href
  {https://doi.org/10.1023/A:1024424200221} {\bibfield  {journal} {\bibinfo
  {journal} {J. Mater. Sci.}\ }\textbf {\bibinfo {volume} {38}},\ \bibinfo
  {pages} {2825--2833} (\bibinfo {year} {2003})}\BibitemShut {NoStop}%
\bibitem [{\citenamefont {Peters}, \citenamefont {Majumdar},\ and\
  \citenamefont {Jaeger}(2016)}]{peters2016direct}%
  \BibitemOpen
  \bibfield  {author} {\bibinfo {author} {\bibfnamefont {I.~R.}\ \bibnamefont
  {Peters}}, \bibinfo {author} {\bibfnamefont {S.}~\bibnamefont {Majumdar}},\
  and\ \bibinfo {author} {\bibfnamefont {H.~M.}\ \bibnamefont {Jaeger}},\
  }\bibfield  {title} {\enquote {\bibinfo {title} {Direct observation of
  dynamic shear jamming in dense suspensions},}\ }\href
  {https://doi.org/10.1038/nature17167} {\bibfield  {journal} {\bibinfo
  {journal} {Nature}\ }\textbf {\bibinfo {volume} {532}},\ \bibinfo {pages}
  {214--217} (\bibinfo {year} {2016})}\BibitemShut {NoStop}%
\bibitem [{\citenamefont {Mawkhlieng}, \citenamefont {Majumdar},\ and\
  \citenamefont {Laha}(2020)}]{mawkhlieng2020review}%
  \BibitemOpen
  \bibfield  {author} {\bibinfo {author} {\bibfnamefont {U.}~\bibnamefont
  {Mawkhlieng}}, \bibinfo {author} {\bibfnamefont {A.}~\bibnamefont
  {Majumdar}},\ and\ \bibinfo {author} {\bibfnamefont {A.}~\bibnamefont
  {Laha}},\ }\bibfield  {title} {\enquote {\bibinfo {title} {A review of
  fibrous materials for soft body armour applications},}\ }\href
  {https://doi.org/10.1039/C9RA06447H} {\bibfield  {journal} {\bibinfo
  {journal} {RSC Adv.}\ }\textbf {\bibinfo {volume} {10}},\ \bibinfo {pages}
  {1066--1086} (\bibinfo {year} {2020})}\BibitemShut {NoStop}%
\bibitem [{\citenamefont {Cai}\ \emph {et~al.}(2021)\citenamefont {Cai},
  \citenamefont {Chen}, \citenamefont {Zhang}, \citenamefont {Wang},\ and\
  \citenamefont {Zhang}}]{cai2021impact}%
  \BibitemOpen
  \bibfield  {author} {\bibinfo {author} {\bibfnamefont {W.}~\bibnamefont
  {Cai}}, \bibinfo {author} {\bibfnamefont {S.}~\bibnamefont {Chen}}, \bibinfo
  {author} {\bibfnamefont {R.}~\bibnamefont {Zhang}}, \bibinfo {author}
  {\bibfnamefont {X.}~\bibnamefont {Wang}},\ and\ \bibinfo {author}
  {\bibfnamefont {X.}~\bibnamefont {Zhang}},\ }\bibfield  {title} {\enquote
  {\bibinfo {title} {Impact-resistant membranes from electrospun fibers with a
  shear-thickening core},}\ }\href
  {https://doi.org/10.1016/j.matchemphys.2021.125478} {\bibfield  {journal}
  {\bibinfo  {journal} {Mater. Chem. Phys.}\ }\textbf {\bibinfo {volume}
  {277}},\ \bibinfo {pages} {125478} (\bibinfo {year} {2021})}\BibitemShut
  {NoStop}%
\bibitem [{\citenamefont {Caglayan}\ \emph {et~al.}(2020)\citenamefont
  {Caglayan}, \citenamefont {Osken}, \citenamefont {Ataalp}, \citenamefont
  {Turkmen},\ and\ \citenamefont {Cebeci}}]{caglayan2020impact}%
  \BibitemOpen
  \bibfield  {author} {\bibinfo {author} {\bibfnamefont {C.}~\bibnamefont
  {Caglayan}}, \bibinfo {author} {\bibfnamefont {I.}~\bibnamefont {Osken}},
  \bibinfo {author} {\bibfnamefont {A.}~\bibnamefont {Ataalp}}, \bibinfo
  {author} {\bibfnamefont {H.~S.}\ \bibnamefont {Turkmen}},\ and\ \bibinfo
  {author} {\bibfnamefont {H.}~\bibnamefont {Cebeci}},\ }\bibfield  {title}
  {\enquote {\bibinfo {title} {Impact response of shear thickening fluid filled
  polyurethane foam core sandwich composites},}\ }\href
  {https://doi.org/10.1016/j.compstruct.2020.112171} {\bibfield  {journal}
  {\bibinfo  {journal} {Compos. Struct.}\ }\textbf {\bibinfo {volume} {243}},\
  \bibinfo {pages} {112171} (\bibinfo {year} {2020})}\BibitemShut {NoStop}%
\bibitem [{\citenamefont {Fischer}\ \emph {et~al.}(2006)\citenamefont
  {Fischer}, \citenamefont {Braun}, \citenamefont {Bourban}, \citenamefont
  {Michaud}, \citenamefont {Plummer},\ and\ \citenamefont
  {M{\aa}nson}}]{fischer2006dynamic}%
  \BibitemOpen
  \bibfield  {author} {\bibinfo {author} {\bibfnamefont {C.}~\bibnamefont
  {Fischer}}, \bibinfo {author} {\bibfnamefont {S.}~\bibnamefont {Braun}},
  \bibinfo {author} {\bibfnamefont {P.}~\bibnamefont {Bourban}}, \bibinfo
  {author} {\bibfnamefont {V.}~\bibnamefont {Michaud}}, \bibinfo {author}
  {\bibfnamefont {C.}~\bibnamefont {Plummer}},\ and\ \bibinfo {author}
  {\bibfnamefont {J.~E.}\ \bibnamefont {M{\aa}nson}},\ }\bibfield  {title}
  {\enquote {\bibinfo {title} {Dynamic properties of sandwich structures with
  integrated shear-thickening fluids},}\ }\href
  {https://doi.org/10.1088/0964-1726/15/5/036} {\bibfield  {journal} {\bibinfo
  {journal} {Smart Mater. Struct.}\ }\textbf {\bibinfo {volume} {15}},\
  \bibinfo {pages} {1467} (\bibinfo {year} {2006})}\BibitemShut {NoStop}%
\bibitem [{\citenamefont {Pinto}\ and\ \citenamefont
  {Meo}(2017)}]{pinto2017design}%
  \BibitemOpen
  \bibfield  {author} {\bibinfo {author} {\bibfnamefont {F.}~\bibnamefont
  {Pinto}}\ and\ \bibinfo {author} {\bibfnamefont {M.}~\bibnamefont {Meo}},\
  }\bibfield  {title} {\enquote {\bibinfo {title} {Design and manufacturing of
  a novel shear thickening fluid composite ({STFC}) with enhanced out-of-plane
  properties and damage suppression},}\ }\href
  {https://doi.org/10.1007/s10443-016-9532-1} {\bibfield  {journal} {\bibinfo
  {journal} {Appl. Compos. Mater.}\ }\textbf {\bibinfo {volume} {24}},\
  \bibinfo {pages} {643--660} (\bibinfo {year} {2017})}\BibitemShut {NoStop}%
\bibitem [{\citenamefont {Myronidis}\ \emph {et~al.}(2022)\citenamefont
  {Myronidis}, \citenamefont {Thielke}, \citenamefont {Kope{\'c}},
  \citenamefont {Meo},\ and\ \citenamefont
  {Pinto}}]{myronidis2022polyborosiloxane}%
  \BibitemOpen
  \bibfield  {author} {\bibinfo {author} {\bibfnamefont {K.}~\bibnamefont
  {Myronidis}}, \bibinfo {author} {\bibfnamefont {M.}~\bibnamefont {Thielke}},
  \bibinfo {author} {\bibfnamefont {M.}~\bibnamefont {Kope{\'c}}}, \bibinfo
  {author} {\bibfnamefont {M.}~\bibnamefont {Meo}},\ and\ \bibinfo {author}
  {\bibfnamefont {F.}~\bibnamefont {Pinto}},\ }\bibfield  {title} {\enquote
  {\bibinfo {title} {Polyborosiloxane-based, dynamic shear stiffening
  multilayer coating for the protection of composite laminates under low
  velocity impact},}\ }\href
  {https://doi.org/10.1016/j.compscitech.2022.109395} {\bibfield  {journal}
  {\bibinfo  {journal} {Compos. Sci. Technol.}\ }\textbf {\bibinfo {volume}
  {222}},\ \bibinfo {pages} {109395} (\bibinfo {year} {2022})}\BibitemShut
  {NoStop}%
\bibitem [{\citenamefont {Hou}, \citenamefont {Wang},\ and\ \citenamefont
  {Layton}(2012)}]{hou2012numerical}%
  \BibitemOpen
  \bibfield  {author} {\bibinfo {author} {\bibfnamefont {G.}~\bibnamefont
  {Hou}}, \bibinfo {author} {\bibfnamefont {J.}~\bibnamefont {Wang}},\ and\
  \bibinfo {author} {\bibfnamefont {A.}~\bibnamefont {Layton}},\ }\bibfield
  {title} {\enquote {\bibinfo {title} {Numerical methods for fluid-structure
  interaction—a review},}\ }\href
  {https://doi.org/10.4208/cicp.291210.290411s} {\bibfield  {journal} {\bibinfo
   {journal} {Commun. Comp. Phys.}\ }\textbf {\bibinfo {volume} {12}},\
  \bibinfo {pages} {337--377} (\bibinfo {year} {2012})}\BibitemShut {NoStop}%
\bibitem [{\citenamefont {Janela}, \citenamefont {Moura},\ and\ \citenamefont
  {Sequeira}(2010)}]{janela20103d}%
  \BibitemOpen
  \bibfield  {author} {\bibinfo {author} {\bibfnamefont {J.}~\bibnamefont
  {Janela}}, \bibinfo {author} {\bibfnamefont {A.}~\bibnamefont {Moura}},\ and\
  \bibinfo {author} {\bibfnamefont {A.}~\bibnamefont {Sequeira}},\ }\bibfield
  {title} {\enquote {\bibinfo {title} {A {3D} non-{N}ewtonian fluid--structure
  interaction model for blood flow in arteries},}\ }\href
  {https://doi.org/10.1016/j.cam.2010.01.032} {\bibfield  {journal} {\bibinfo
  {journal} {J. Comput. Appl. Math.}\ }\textbf {\bibinfo {volume} {234}},\
  \bibinfo {pages} {2783--2791} (\bibinfo {year} {2010})}\BibitemShut {NoStop}%
\bibitem [{\citenamefont {Anand}, \citenamefont {David J.~R.},\ and\
  \citenamefont {Christov}(2019)}]{anand2019non}%
  \BibitemOpen
  \bibfield  {author} {\bibinfo {author} {\bibfnamefont {V.}~\bibnamefont
  {Anand}}, \bibinfo {author} {\bibfnamefont {J.}~\bibnamefont {David J.~R.}},\
  and\ \bibinfo {author} {\bibfnamefont {I.~C.}\ \bibnamefont {Christov}},\
  }\bibfield  {title} {\enquote {\bibinfo {title} {Non-{N}ewtonian
  fluid--structure interactions: Static response of a microchannel due to
  internal flow of a power-law fluid},}\ }\href
  {https://doi.org/10.1016/j.jnnfm.2018.12.008} {\bibfield  {journal} {\bibinfo
   {journal} {J. Non-Newton Fluid Mech.}\ }\textbf {\bibinfo {volume} {264}},\
  \bibinfo {pages} {62--72} (\bibinfo {year} {2019})}\BibitemShut {NoStop}%
\bibitem [{\citenamefont {Zhu}\ \emph {et~al.}(2017)\citenamefont {Zhu},
  \citenamefont {Yu}, \citenamefont {Liu}, \citenamefont {Cheng},\ and\
  \citenamefont {Lu}}]{zhu2017deformable}%
  \BibitemOpen
  \bibfield  {author} {\bibinfo {author} {\bibfnamefont {L.}~\bibnamefont
  {Zhu}}, \bibinfo {author} {\bibfnamefont {X.}~\bibnamefont {Yu}}, \bibinfo
  {author} {\bibfnamefont {N.}~\bibnamefont {Liu}}, \bibinfo {author}
  {\bibfnamefont {Y.}~\bibnamefont {Cheng}},\ and\ \bibinfo {author}
  {\bibfnamefont {X.}~\bibnamefont {Lu}},\ }\bibfield  {title} {\enquote
  {\bibinfo {title} {A deformable plate interacting with a non-{N}ewtonian
  fluid in three dimensions},}\ }\href {https://doi.org/10.1063/1.4996040}
  {\bibfield  {journal} {\bibinfo  {journal} {Phys. Fluids}\ }\textbf {\bibinfo
  {volume} {29}},\ \bibinfo {pages} {083101} (\bibinfo {year}
  {2017})}\BibitemShut {NoStop}%
\bibitem [{\citenamefont {Krapez}\ \emph {et~al.}(2022)\citenamefont {Krapez},
  \citenamefont {Gauthier}, \citenamefont {Boitte}, \citenamefont {Aubrun},
  \citenamefont {Joanny},\ and\ \citenamefont {Colin}}]{krapez2022spreading}%
  \BibitemOpen
  \bibfield  {author} {\bibinfo {author} {\bibfnamefont {M.}~\bibnamefont
  {Krapez}}, \bibinfo {author} {\bibfnamefont {A.}~\bibnamefont {Gauthier}},
  \bibinfo {author} {\bibfnamefont {J.-B.}\ \bibnamefont {Boitte}}, \bibinfo
  {author} {\bibfnamefont {O.}~\bibnamefont {Aubrun}}, \bibinfo {author}
  {\bibfnamefont {J.-F.}\ \bibnamefont {Joanny}},\ and\ \bibinfo {author}
  {\bibfnamefont {A.}~\bibnamefont {Colin}},\ }\bibfield  {title} {\enquote
  {\bibinfo {title} {Spreading of complex fluids with a soft blade},}\ }\href
  {https://doi.org/10.1103/PhysRevFluids.7.084002} {\bibfield  {journal}
  {\bibinfo  {journal} {Phys. Rev. Fluids}\ }\textbf {\bibinfo {volume}
  {7}},\ \bibinfo {pages} {084002} (\bibinfo {year} {2022})}\BibitemShut
  {NoStop}%
\bibitem [{\citenamefont {Bauchau}\ and\ \citenamefont
  {Craig}(2009)}]{bauchau2009euler}%
  \BibitemOpen
  \bibfield  {author} {\bibinfo {author} {\bibfnamefont {O.~A.}\ \bibnamefont
  {Bauchau}}\ and\ \bibinfo {author} {\bibfnamefont {J.~I.}\ \bibnamefont
  {Craig}},\ }\enquote {\bibinfo {title} {{E}uler-{B}ernoulli beam theory},}\
  in\ \href {https://doi.org/10.1007/978-90-481-2516-6_5} {\emph {\bibinfo
  {booktitle} {Structural Analysis}}},\ \bibinfo {editor} {edited by\ \bibinfo
  {editor} {\bibfnamefont {O.~A.}\ \bibnamefont {Bauchau}}\ and\ \bibinfo
  {editor} {\bibfnamefont {J.~I.}\ \bibnamefont {Craig}}}\ (\bibinfo
  {publisher} {Springer Netherlands},\ \bibinfo {address} {Dordrecht},\
  \bibinfo {year} {2009})\ pp.\ \bibinfo {pages} {173--221}\BibitemShut
  {NoStop}%
\bibitem [{\citenamefont {Hamrock}, \citenamefont {Schmid},\ and\ \citenamefont
  {Jacobson}(2004)}]{hamrock2004fundamentals}%
  \BibitemOpen
  \bibfield  {author} {\bibinfo {author} {\bibfnamefont {B.~J.}\ \bibnamefont
  {Hamrock}}, \bibinfo {author} {\bibfnamefont {S.~R.}\ \bibnamefont
  {Schmid}},\ and\ \bibinfo {author} {\bibfnamefont {B.~O.}\ \bibnamefont
  {Jacobson}},\ }\href {https://doi.org/10.1201/9780203021187} {\emph {\bibinfo
  {title} {Fundamentals of Fluid Film Lubrication}}}\ (\bibinfo  {publisher}
  {CRC Press},\ \bibinfo {address} {Boca Raton},\ \bibinfo {year}
  {2004})\BibitemShut {NoStop}%
\bibitem [{\citenamefont {Zhu}(2019)}]{zhu2019ib}%
  \BibitemOpen
  \bibfield  {author} {\bibinfo {author} {\bibfnamefont {L.}~\bibnamefont
  {Zhu}},\ }\bibfield  {title} {\enquote {\bibinfo {title} {An {IB} method for
  non-{N}ewtonian-fluid flexible-structure interactions in three-dimensions},}\
  }\href@noop {} {\bibfield  {journal} {\bibinfo  {journal} {Comput. Model Eng.
  Sci.}\ }\textbf {\bibinfo {volume} {119}},\ \bibinfo {pages} {125--143}
  (\bibinfo {year} {2019})}\BibitemShut {NoStop}%
\bibitem [{\citenamefont {Xu}(2016)}]{xu2016numerical}%
  \BibitemOpen
  \bibfield  {author} {\bibinfo {author} {\bibfnamefont {S.}~\bibnamefont
  {Xu}},\ }\emph {\bibinfo {title} {Numerical study for non-{N}ewtonian
  fluid-structure interaction problems}},\ \href@noop {} {Ph.D. thesis},\
  \bibinfo  {school} {Clemson University} (\bibinfo {year} {2016})\BibitemShut
  {NoStop}%
\bibitem [{Note1()}]{Note1}%
  \BibitemOpen
  \bibinfo {note} {Initial contact is not accurately described, but for large
  deformations ($h_0 \to 0$) this can be neglected.}\BibitemShut {Stop}%
\bibitem [{\citenamefont {Gibson}\ \emph {et~al.}(1998)\citenamefont {Gibson},
  \citenamefont {Kotsikos}, \citenamefont {Bland},\ and\ \citenamefont
  {Toll}}]{gibson1998squeeze}%
  \BibitemOpen
  \bibfield  {author} {\bibinfo {author} {\bibfnamefont {A.}~\bibnamefont
  {Gibson}}, \bibinfo {author} {\bibfnamefont {G.}~\bibnamefont {Kotsikos}},
  \bibinfo {author} {\bibfnamefont {J.}~\bibnamefont {Bland}},\ and\ \bibinfo
  {author} {\bibfnamefont {S.}~\bibnamefont {Toll}},\ }\bibfield  {title}
  {\enquote {\bibinfo {title} {Squeeze flow},}\ }in\ \href
  {https://doi.org/10.1007/978-94-011-4934-1_18} {\emph {\bibinfo {booktitle}
  {Rheological Measurement}}}\ (\bibinfo  {publisher} {Springer},\ \bibinfo
  {year} {1998})\ pp.\ \bibinfo {pages} {550--592}\BibitemShut {NoStop}%
\bibitem [{Note2()}]{Note2}%
  \BibitemOpen
  \bibinfo {note} {Larger $w_\protect \mathrm {eff}$ increases $Q$ and
  $\partial _{y}p \propto w_\protect \mathrm {eff}$, such that $p \propto
  w_\protect \mathrm {eff}^2$ and $F \propto w_\protect \mathrm {eff}^3$. The
  bending moment in the plate $\propto w_\protect \mathrm {eff}^4$, the angular
  deflection $\propto w_\protect \mathrm {eff}^5$ and, ultimately, $\Delta h
  \propto w_\protect \mathrm {eff}^6$.}\BibitemShut {Stop}%
\bibitem [{\citenamefont {Macosko}(1994)}]{macosko1994rheology}%
  \BibitemOpen
  \bibfield  {author} {\bibinfo {author} {\bibfnamefont {C.~W.}\ \bibnamefont
  {Macosko}},\ }\href@noop {} {\emph {\bibinfo {title} {Rheology: Principles,
  Measurements and Applications}}},\ Advances in interfacial engineering\
  (\bibinfo  {publisher} {Wiley-VCH},\ \bibinfo {year} {1994})\BibitemShut
  {NoStop}%
\bibitem [{\citenamefont {Meeten}(2005)}]{meeten2005flow}%
  \BibitemOpen
  \bibfield  {author} {\bibinfo {author} {\bibfnamefont {G.~H.}\ \bibnamefont
  {Meeten}},\ }\bibfield  {title} {\enquote {\bibinfo {title} {Flow of soft
  solids squeezed between planar and spherical surfaces},}\ }\href
  {https://doi.org/10.1007/s00397-005-0437-4} {\bibfield  {journal} {\bibinfo
  {journal} {Rheol. Acta.}\ }\textbf {\bibinfo {volume} {44}},\ \bibinfo
  {pages} {563--572} (\bibinfo {year} {2005})}\BibitemShut {NoStop}%
\bibitem [{\citenamefont {Di~Giuseppe}\ \emph {et~al.}(2012)\citenamefont
  {Di~Giuseppe}, \citenamefont {Davaille}, \citenamefont {Mittelstaedt},\ and\
  \citenamefont {Fran{\c{c}}ois}}]{di2012rheological}%
  \BibitemOpen
  \bibfield  {author} {\bibinfo {author} {\bibfnamefont {E.}~\bibnamefont
  {Di~Giuseppe}}, \bibinfo {author} {\bibfnamefont {A.}~\bibnamefont
  {Davaille}}, \bibinfo {author} {\bibfnamefont {E.}~\bibnamefont
  {Mittelstaedt}},\ and\ \bibinfo {author} {\bibfnamefont {M.}~\bibnamefont
  {Fran{\c{c}}ois}},\ }\bibfield  {title} {\enquote {\bibinfo {title}
  {Rheological and mechanical properties of silica colloids: from {N}ewtonian
  liquid to brittle behaviour},}\ }\href
  {https://doi.org/10.1007/s00397-011-0611-9} {\bibfield  {journal} {\bibinfo
  {journal} {Rheol. Acta}\ }\textbf {\bibinfo {volume} {51}},\ \bibinfo {pages}
  {451--465} (\bibinfo {year} {2012})}\BibitemShut {NoStop}%
\bibitem [{\citenamefont {Lobry}\ \emph {et~al.}(2019)\citenamefont {Lobry},
  \citenamefont {Lemaire}, \citenamefont {Blanc}, \citenamefont {Gallier},\
  and\ \citenamefont {Peters}}]{lobry2019shear}%
  \BibitemOpen
  \bibfield  {author} {\bibinfo {author} {\bibfnamefont {L.}~\bibnamefont
  {Lobry}}, \bibinfo {author} {\bibfnamefont {E.}~\bibnamefont {Lemaire}},
  \bibinfo {author} {\bibfnamefont {F.}~\bibnamefont {Blanc}}, \bibinfo
  {author} {\bibfnamefont {S.}~\bibnamefont {Gallier}},\ and\ \bibinfo {author}
  {\bibfnamefont {F.}~\bibnamefont {Peters}},\ }\bibfield  {title} {\enquote
  {\bibinfo {title} {Shear thinning in non-{B}rownian suspensions explained by
  variable friction between particles},}\ }\href
  {https://doi.org/10.1017/jfm.2018.881} {\bibfield  {journal} {\bibinfo
  {journal} {J. Fluid Mech.}\ }\textbf {\bibinfo {volume} {860}},\ \bibinfo
  {pages} {682--710} (\bibinfo {year} {2019})}\BibitemShut {NoStop}%
\bibitem [{\citenamefont {Le}\ \emph {et~al.}(2023)\citenamefont {Le},
  \citenamefont {Izzet}, \citenamefont {Ovarlez},\ and\ \citenamefont
  {Colin}}]{nguyenle2022solvents}%
  \BibitemOpen
  \bibfield  {author} {\bibinfo {author} {\bibfnamefont {A.~V.~N.}\
  \bibnamefont {Le}}, \bibinfo {author} {\bibfnamefont {A.}~\bibnamefont
  {Izzet}}, \bibinfo {author} {\bibfnamefont {G.}~\bibnamefont {Ovarlez}},\
  and\ \bibinfo {author} {\bibfnamefont {A.}~\bibnamefont {Colin}},\ }\bibfield
   {title} {\enquote {\bibinfo {title} {Solvents govern rheology and jamming of
  polymeric bead suspensions},}\ }\href
  {https://doi.org/10.1016/j.jcis.2022.09.074} {\bibfield  {journal} {\bibinfo
  {journal} {Journal of Colloid and Interface Science}\ }\textbf {\bibinfo
  {volume} {629}},\ \bibinfo {pages} {438--450} (\bibinfo {year}
  {2023})}\BibitemShut {NoStop}%
\bibitem [{\citenamefont {Ryder}\ and\ \citenamefont
  {Yeomans}(2006)}]{ryder2006shear}%
  \BibitemOpen
  \bibfield  {author} {\bibinfo {author} {\bibfnamefont {J.~F.}\ \bibnamefont
  {Ryder}}\ and\ \bibinfo {author} {\bibfnamefont {J.~M.}\ \bibnamefont
  {Yeomans}},\ }\bibfield  {title} {\enquote {\bibinfo {title} {Shear thinning
  in dilute polymer solutions},}\ }\href {https://doi.org/10.1063/1.2387948}
  {\bibfield  {journal} {\bibinfo  {journal} {J. Chem. Phys.}\ }\textbf
  {\bibinfo {volume} {125}},\ \bibinfo {pages} {194906} (\bibinfo {year}
  {2006})}\BibitemShut {NoStop}%
\bibitem [{\citenamefont {Laun}, \citenamefont {Rady},\ and\ \citenamefont
  {Hassager}(1999)}]{laun1999analytical}%
  \BibitemOpen
  \bibfield  {author} {\bibinfo {author} {\bibfnamefont {H.}~\bibnamefont
  {Laun}}, \bibinfo {author} {\bibfnamefont {M.}~\bibnamefont {Rady}},\ and\
  \bibinfo {author} {\bibfnamefont {O.}~\bibnamefont {Hassager}},\ }\bibfield
  {title} {\enquote {\bibinfo {title} {Analytical solutions for squeeze flow
  with partial wall slip},}\ }\href
  {https://doi.org/10.1016/S0377-0257(98)00083-4} {\bibfield  {journal}
  {\bibinfo  {journal} {J. Non-Newton Fluid Mech.}\ }\textbf {\bibinfo {volume}
  {81}},\ \bibinfo {pages} {1--15} (\bibinfo {year} {1999})}\BibitemShut
  {NoStop}%
\bibitem [{\citenamefont {Royer}, \citenamefont {Blair},\ and\ \citenamefont
  {Hudson}(2016)}]{royer2016rheological}%
  \BibitemOpen
  \bibfield  {author} {\bibinfo {author} {\bibfnamefont {J.~R.}\ \bibnamefont
  {Royer}}, \bibinfo {author} {\bibfnamefont {D.~L.}\ \bibnamefont {Blair}},\
  and\ \bibinfo {author} {\bibfnamefont {S.~D.}\ \bibnamefont {Hudson}},\
  }\bibfield  {title} {\enquote {\bibinfo {title} {Rheological signature of
  frictional interactions in shear thickening suspensions},}\ }\href
  {https://doi.org/10.1103/PhysRevLett.116.188301} {\bibfield  {journal}
  {\bibinfo  {journal} {Phys. Rev. Lett.}\ }\textbf {\bibinfo {volume} {116}},\
  \bibinfo {pages} {188301} (\bibinfo {year} {2016})}\BibitemShut {NoStop}%
\bibitem [{\citenamefont {Richards}\ \emph {et~al.}(2019)\citenamefont
  {Richards}, \citenamefont {Royer}, \citenamefont {Liebchen}, \citenamefont
  {Guy},\ and\ \citenamefont {Poon}}]{richards2019competing}%
  \BibitemOpen
  \bibfield  {author} {\bibinfo {author} {\bibfnamefont {J.~A.}\ \bibnamefont
  {Richards}}, \bibinfo {author} {\bibfnamefont {J.~R.}\ \bibnamefont {Royer}},
  \bibinfo {author} {\bibfnamefont {B.}~\bibnamefont {Liebchen}}, \bibinfo
  {author} {\bibfnamefont {B.~M.}\ \bibnamefont {Guy}},\ and\ \bibinfo {author}
  {\bibfnamefont {W.~C.~K.}\ \bibnamefont {Poon}},\ }\bibfield  {title}
  {\enquote {\bibinfo {title} {Competing timescales lead to oscillations in
  shear-thickening suspensions},}\ }\href
  {https://doi.org/10.1103/PhysRevLett.123.038004} {\bibfield  {journal}
  {\bibinfo  {journal} {Phys. Rev. Lett.}\ }\textbf {\bibinfo {volume} {123}},\
  \bibinfo {pages} {038004} (\bibinfo {year} {2019})}\BibitemShut {NoStop}%
\bibitem [{\citenamefont {Cheal}\ and\ \citenamefont
  {Ness}(2018)}]{cheal2018rheology}%
  \BibitemOpen
  \bibfield  {author} {\bibinfo {author} {\bibfnamefont {O.}~\bibnamefont
  {Cheal}}\ and\ \bibinfo {author} {\bibfnamefont {C.}~\bibnamefont {Ness}},\
  }\bibfield  {title} {\enquote {\bibinfo {title} {Rheology of dense granular
  suspensions under extensional flow},}\ }\href
  {https://doi.org/10.1122/1.5004007} {\bibfield  {journal} {\bibinfo
  {journal} {J. Rheol.}\ }\textbf {\bibinfo {volume} {62}},\ \bibinfo {pages}
  {501--512} (\bibinfo {year} {2018})}\BibitemShut {NoStop}%
\bibitem [{\citenamefont {Cyriac}\ \emph {et~al.}(2022)\citenamefont {Cyriac},
  \citenamefont {Xin~Yi}, \citenamefont {Chow},\ and\ \citenamefont
  {Macbeath}}]{cyriac2022tactile}%
  \BibitemOpen
  \bibfield  {author} {\bibinfo {author} {\bibfnamefont {F.}~\bibnamefont
  {Cyriac}}, \bibinfo {author} {\bibfnamefont {T.}~\bibnamefont {Xin~Yi}},
  \bibinfo {author} {\bibfnamefont {P.~S.}\ \bibnamefont {Chow}},\ and\
  \bibinfo {author} {\bibfnamefont {C.}~\bibnamefont {Macbeath}},\ }\bibfield
  {title} {\enquote {\bibinfo {title} {Tactile friction and rheological studies
  to objectify sensory properties of topical formulations},}\ }\href
  {https://doi.org/10.1122/8.0000341} {\bibfield  {journal} {\bibinfo
  {journal} {J. Rheol.}\ }\textbf {\bibinfo {volume} {66}},\ \bibinfo {pages}
  {305--326} (\bibinfo {year} {2022})}\BibitemShut {NoStop}%
\bibitem [{\citenamefont {Rodrigues}\ \emph {et~al.}(2021)\citenamefont
  {Rodrigues}, \citenamefont {Shewan}, \citenamefont {Xu}, \citenamefont
  {Selway},\ and\ \citenamefont {Stokes}}]{rodrigues2021frictional}%
  \BibitemOpen
  \bibfield  {author} {\bibinfo {author} {\bibfnamefont {S.~A.}\ \bibnamefont
  {Rodrigues}}, \bibinfo {author} {\bibfnamefont {H.~M.}\ \bibnamefont
  {Shewan}}, \bibinfo {author} {\bibfnamefont {Y.}~\bibnamefont {Xu}}, \bibinfo
  {author} {\bibfnamefont {N.}~\bibnamefont {Selway}},\ and\ \bibinfo {author}
  {\bibfnamefont {J.~R.}\ \bibnamefont {Stokes}},\ }\bibfield  {title}
  {\enquote {\bibinfo {title} {Frictional behaviour of molten chocolate as a
  function of fat content},}\ }\href {https://doi.org/10.1039/D0FO03378B}
  {\bibfield  {journal} {\bibinfo  {journal} {Food \& Function}\ }\textbf
  {\bibinfo {volume} {12}},\ \bibinfo {pages} {2457--2467} (\bibinfo {year}
  {2021})}\BibitemShut {NoStop}%
\bibitem [{\citenamefont {Ibaseta}\ and\ \citenamefont
  {Biscans}(2010)}]{ibaseta2010fractal}%
  \BibitemOpen
  \bibfield  {author} {\bibinfo {author} {\bibfnamefont {N.}~\bibnamefont
  {Ibaseta}}\ and\ \bibinfo {author} {\bibfnamefont {B.}~\bibnamefont
  {Biscans}},\ }\bibfield  {title} {\enquote {\bibinfo {title} {Fractal
  dimension of fumed silica: Comparison of light scattering and electron
  microscope methods},}\ }\href {https://doi.org/10.1016/j.powtec.2010.05.010}
  {\bibfield  {journal} {\bibinfo  {journal} {Powder Technol.}\ }\textbf
  {\bibinfo {volume} {203}},\ \bibinfo {pages} {206--210} (\bibinfo {year}
  {2010})}\BibitemShut {NoStop}%
\bibitem [{\citenamefont {Kamaly}, \citenamefont {Tarleton},\ and\
  \citenamefont {{\"O}zcan-Ta{\c{s}}k{\i}n}(2017)}]{kamaly2017dispersion}%
  \BibitemOpen
  \bibfield  {author} {\bibinfo {author} {\bibfnamefont {S.~W.}\ \bibnamefont
  {Kamaly}}, \bibinfo {author} {\bibfnamefont {A.~C.}\ \bibnamefont
  {Tarleton}},\ and\ \bibinfo {author} {\bibfnamefont {N.~G.}\ \bibnamefont
  {{\"O}zcan-Ta{\c{s}}k{\i}n}},\ }\bibfield  {title} {\enquote {\bibinfo
  {title} {Dispersion of clusters of nanoscale silica particles using batch
  rotor-stators},}\ }\href {https://doi.org/10.1016/j.apt.2017.06.017}
  {\bibfield  {journal} {\bibinfo  {journal} {Adv. Powder Technol.}\ }\textbf
  {\bibinfo {volume} {28}},\ \bibinfo {pages} {2357--2365} (\bibinfo {year}
  {2017})}\BibitemShut {NoStop}%
\bibitem [{\citenamefont {Blanco}\ \emph {et~al.}(2019)\citenamefont {Blanco},
  \citenamefont {Hodgson}, \citenamefont {Hermes}, \citenamefont {Besseling},
  \citenamefont {Hunter}, \citenamefont {Chaikin}, \citenamefont {Cates},
  \citenamefont {Van~Damme},\ and\ \citenamefont {Poon}}]{blanco2019conching}%
  \BibitemOpen
  \bibfield  {author} {\bibinfo {author} {\bibfnamefont {E.}~\bibnamefont
  {Blanco}}, \bibinfo {author} {\bibfnamefont {D.~J.}\ \bibnamefont {Hodgson}},
  \bibinfo {author} {\bibfnamefont {M.}~\bibnamefont {Hermes}}, \bibinfo
  {author} {\bibfnamefont {R.}~\bibnamefont {Besseling}}, \bibinfo {author}
  {\bibfnamefont {G.~L.}\ \bibnamefont {Hunter}}, \bibinfo {author}
  {\bibfnamefont {P.~M.}\ \bibnamefont {Chaikin}}, \bibinfo {author}
  {\bibfnamefont {M.~E.}\ \bibnamefont {Cates}}, \bibinfo {author}
  {\bibfnamefont {I.}~\bibnamefont {Van~Damme}},\ and\ \bibinfo {author}
  {\bibfnamefont {W.~C.~K.}\ \bibnamefont {Poon}},\ }\bibfield  {title}
  {\enquote {\bibinfo {title} {Conching chocolate is a prototypical transition
  from frictionally jammed solid to flowable suspension with maximal solid
  content},}\ }\href {https://doi.org/10.1073/pnas.1901858116} {\bibfield
  {journal} {\bibinfo  {journal} {Proc. Natl. Acad. Sci. U.S.A.}\ }\textbf
  {\bibinfo {volume} {116}},\ \bibinfo {pages} {10303--10308} (\bibinfo {year}
  {2019})}\BibitemShut {NoStop}%
\bibitem [{\citenamefont {Barnes}(1989)}]{barnes1989shear}%
  \BibitemOpen
  \bibfield  {author} {\bibinfo {author} {\bibfnamefont {H.~A.}\ \bibnamefont
  {Barnes}},\ }\bibfield  {title} {\enquote {\bibinfo {title} {Shear-thickening
  (“dilatancy”) in suspensions of nonaggregating solid particles dispersed
  in {N}ewtonian liquids},}\ }\href {https://doi.org/10.1122/1.550017}
  {\bibfield  {journal} {\bibinfo  {journal} {J. Rheol.}\ }\textbf {\bibinfo
  {volume} {33}},\ \bibinfo {pages} {329--366} (\bibinfo {year}
  {1989})}\BibitemShut {NoStop}%
\bibitem [{\citenamefont {van~der Naald}\ \emph {et~al.}(2021)\citenamefont
  {van~der Naald}, \citenamefont {Zhao}, \citenamefont {Jackson},\ and\
  \citenamefont {Jaeger}}]{van2021role}%
  \BibitemOpen
  \bibfield  {author} {\bibinfo {author} {\bibfnamefont {M.}~\bibnamefont
  {van~der Naald}}, \bibinfo {author} {\bibfnamefont {L.}~\bibnamefont {Zhao}},
  \bibinfo {author} {\bibfnamefont {G.~L.}\ \bibnamefont {Jackson}},\ and\
  \bibinfo {author} {\bibfnamefont {H.~M.}\ \bibnamefont {Jaeger}},\ }\bibfield
   {title} {\enquote {\bibinfo {title} {The role of solvent molecular weight in
  shear thickening and shear jamming},}\ }\href
  {https://doi.org/10.1039/D0SM01350A} {\bibfield  {journal} {\bibinfo
  {journal} {Soft Matter}\ }\textbf {\bibinfo {volume} {17}},\ \bibinfo {pages}
  {3144--3152} (\bibinfo {year} {2021})}\BibitemShut {NoStop}%
\bibitem [{\citenamefont {Lin}\ \emph {et~al.}(2015)\citenamefont {Lin},
  \citenamefont {Guy}, \citenamefont {Hermes}, \citenamefont {Ness},
  \citenamefont {Sun}, \citenamefont {Poon},\ and\ \citenamefont
  {Cohen}}]{lin2015hydrodynamic}%
  \BibitemOpen
  \bibfield  {author} {\bibinfo {author} {\bibfnamefont {N.~Y.~C.}\
  \bibnamefont {Lin}}, \bibinfo {author} {\bibfnamefont {B.~M.}\ \bibnamefont
  {Guy}}, \bibinfo {author} {\bibfnamefont {M.}~\bibnamefont {Hermes}},
  \bibinfo {author} {\bibfnamefont {C.}~\bibnamefont {Ness}}, \bibinfo {author}
  {\bibfnamefont {J.}~\bibnamefont {Sun}}, \bibinfo {author} {\bibfnamefont
  {W.~C.~K.}\ \bibnamefont {Poon}},\ and\ \bibinfo {author} {\bibfnamefont
  {I.}~\bibnamefont {Cohen}},\ }\bibfield  {title} {\enquote {\bibinfo {title}
  {Hydrodynamic and contact contributions to continuous shear thickening in
  colloidal suspensions},}\ }\href
  {https://doi.org/10.1103/PhysRevLett.115.228304} {\bibfield  {journal}
  {\bibinfo  {journal} {Phys. Rev. Lett.}\ }\textbf {\bibinfo {volume} {115}},\
  \bibinfo {pages} {228304} (\bibinfo {year} {2015})}\BibitemShut {NoStop}%
\bibitem [{\citenamefont {Guy}\ \emph {et~al.}(2018)\citenamefont {Guy},
  \citenamefont {Richards}, \citenamefont {Hodgson}, \citenamefont {Blanco},\
  and\ \citenamefont {Poon}}]{guy2018constraint}%
  \BibitemOpen
  \bibfield  {author} {\bibinfo {author} {\bibfnamefont {B.~M.}\ \bibnamefont
  {Guy}}, \bibinfo {author} {\bibfnamefont {J.~A.}\ \bibnamefont {Richards}},
  \bibinfo {author} {\bibfnamefont {D.~J.~M.}\ \bibnamefont {Hodgson}},
  \bibinfo {author} {\bibfnamefont {E.}~\bibnamefont {Blanco}},\ and\ \bibinfo
  {author} {\bibfnamefont {W.~C.~K.}\ \bibnamefont {Poon}},\ }\bibfield
  {title} {\enquote {\bibinfo {title} {Constraint-based approach to granular
  dispersion rheology},}\ }\href
  {https://doi.org/10.1103/PhysRevLett.121.128001} {\bibfield  {journal}
  {\bibinfo  {journal} {Phys. Rev. Lett.}\ }\textbf {\bibinfo {volume} {121}},\
  \bibinfo {pages} {128001} (\bibinfo {year} {2018})}\BibitemShut {NoStop}%
\bibitem [{\citenamefont {Singh}\ \emph {et~al.}(2020)\citenamefont {Singh},
  \citenamefont {Ness}, \citenamefont {Seto}, \citenamefont {de~Pablo},\ and\
  \citenamefont {Jaeger}}]{singh2020shear}%
  \BibitemOpen
  \bibfield  {author} {\bibinfo {author} {\bibfnamefont {A.}~\bibnamefont
  {Singh}}, \bibinfo {author} {\bibfnamefont {C.}~\bibnamefont {Ness}},
  \bibinfo {author} {\bibfnamefont {R.}~\bibnamefont {Seto}}, \bibinfo {author}
  {\bibfnamefont {J.~J.}\ \bibnamefont {de~Pablo}},\ and\ \bibinfo {author}
  {\bibfnamefont {H.~M.}\ \bibnamefont {Jaeger}},\ }\bibfield  {title}
  {\enquote {\bibinfo {title} {Shear thickening and jamming of dense
  suspensions: the “roll” of friction},}\ }\href
  {https://doi.org/10.1103/PhysRevLett.124.248005} {\bibfield  {journal}
  {\bibinfo  {journal} {Phys. Rev. Lett.}\ }\textbf {\bibinfo {volume} {124}},\
  \bibinfo {pages} {248005} (\bibinfo {year} {2020})}\BibitemShut {NoStop}%
\bibitem [{\citenamefont {Raghavan}\ \emph {et~al.}(1998)\citenamefont
  {Raghavan}, \citenamefont {Riley}, \citenamefont {Fedkiw},\ and\
  \citenamefont {Khan}}]{raghavan1998composite}%
  \BibitemOpen
  \bibfield  {author} {\bibinfo {author} {\bibfnamefont {S.~R.}\ \bibnamefont
  {Raghavan}}, \bibinfo {author} {\bibfnamefont {M.~W.}\ \bibnamefont {Riley}},
  \bibinfo {author} {\bibfnamefont {P.~S.}\ \bibnamefont {Fedkiw}},\ and\
  \bibinfo {author} {\bibfnamefont {S.~A.}\ \bibnamefont {Khan}},\ }\bibfield
  {title} {\enquote {\bibinfo {title} {Composite polymer electrolytes based on
  poly (ethylene glycol) and hydrophobic fumed silica: dynamic rheology and
  microstructure},}\ }\href {https://doi.org/10.1021/cm970406j} {\bibfield
  {journal} {\bibinfo  {journal} {Chem. Mater.}\ }\textbf {\bibinfo {volume}
  {10}},\ \bibinfo {pages} {244--251} (\bibinfo {year} {1998})}\BibitemShut
  {NoStop}%
\bibitem [{\citenamefont {Richards}, \citenamefont {O’Neill},\ and\
  \citenamefont {Poon}(2021)}]{richards2021turning}%
  \BibitemOpen
  \bibfield  {author} {\bibinfo {author} {\bibfnamefont {J.~A.}\ \bibnamefont
  {Richards}}, \bibinfo {author} {\bibfnamefont {R.~E.}\ \bibnamefont
  {O’Neill}},\ and\ \bibinfo {author} {\bibfnamefont {W.~C.~K.}\ \bibnamefont
  {Poon}},\ }\bibfield  {title} {\enquote {\bibinfo {title} {Turning a
  yield-stress calcite suspension into a shear-thickening one by tuning
  inter-particle friction},}\ }\href
  {https://doi.org/10.1007/s00397-020-01247-z} {\bibfield  {journal} {\bibinfo
  {journal} {Rheol. Acta}\ }\textbf {\bibinfo {volume} {60}},\ \bibinfo {pages}
  {97--106} (\bibinfo {year} {2021})}\BibitemShut {NoStop}%
\bibitem [{\citenamefont {Walls}\ \emph {et~al.}(2003)\citenamefont {Walls},
  \citenamefont {Caines}, \citenamefont {Sanchez},\ and\ \citenamefont
  {Khan}}]{walls2003yield}%
  \BibitemOpen
  \bibfield  {author} {\bibinfo {author} {\bibfnamefont {H.~J.}\ \bibnamefont
  {Walls}}, \bibinfo {author} {\bibfnamefont {S.~B.}\ \bibnamefont {Caines}},
  \bibinfo {author} {\bibfnamefont {A.~M.}\ \bibnamefont {Sanchez}},\ and\
  \bibinfo {author} {\bibfnamefont {S.~A.}\ \bibnamefont {Khan}},\ }\bibfield
  {title} {\enquote {\bibinfo {title} {Yield stress and wall slip phenomena in
  colloidal silica gels},}\ }\href {https://doi.org/10.1122/1.1574023}
  {\bibfield  {journal} {\bibinfo  {journal} {J. Rheol.}\ }\textbf {\bibinfo
  {volume} {47}},\ \bibinfo {pages} {847--868} (\bibinfo {year}
  {2003})}\BibitemShut {NoStop}%
\bibitem [{\citenamefont {Johnston}\ \emph {et~al.}(2014)\citenamefont
  {Johnston}, \citenamefont {McCluskey}, \citenamefont {Tan},\ and\
  \citenamefont {Tracey}}]{johnston2014mechanical}%
  \BibitemOpen
  \bibfield  {author} {\bibinfo {author} {\bibfnamefont {I.~D.}\ \bibnamefont
  {Johnston}}, \bibinfo {author} {\bibfnamefont {D.~K.}\ \bibnamefont
  {McCluskey}}, \bibinfo {author} {\bibfnamefont {C.~K.~L.}\ \bibnamefont
  {Tan}},\ and\ \bibinfo {author} {\bibfnamefont {M.}~\bibnamefont {Tracey}},\
  }\bibfield  {title} {\enquote {\bibinfo {title} {Mechanical characterization
  of bulk {Sylgard} 184 for microfluidics and microengineering},}\ }\href
  {https://doi.org/10.1088/0960-1317/24/3/035017} {\bibfield  {journal}
  {\bibinfo  {journal} {J. Micromech. Microeng.}\ }\textbf {\bibinfo {volume}
  {24}},\ \bibinfo {pages} {035017} (\bibinfo {year} {2014})}\BibitemShut
  {NoStop}%
\bibitem [{\citenamefont {Grandes}\ \emph {et~al.}(2021)\citenamefont
  {Grandes}, \citenamefont {Sakano}, \citenamefont {Rego}, \citenamefont
  {Rebmann}, \citenamefont {Cardoso},\ and\ \citenamefont
  {Pileggi}}]{grandes2021rheological}%
  \BibitemOpen
  \bibfield  {author} {\bibinfo {author} {\bibfnamefont {F.~A.}\ \bibnamefont
  {Grandes}}, \bibinfo {author} {\bibfnamefont {V.~K.}\ \bibnamefont {Sakano}},
  \bibinfo {author} {\bibfnamefont {A.~C.}\ \bibnamefont {Rego}}, \bibinfo
  {author} {\bibfnamefont {M.~S.}\ \bibnamefont {Rebmann}}, \bibinfo {author}
  {\bibfnamefont {F.~A.}\ \bibnamefont {Cardoso}},\ and\ \bibinfo {author}
  {\bibfnamefont {R.~G.}\ \bibnamefont {Pileggi}},\ }\bibfield  {title}
  {\enquote {\bibinfo {title} {Rheological behavior and flow induced
  microstructural changes of cement-based mortars assessed by pressure mapped
  squeeze flow},}\ }\href {https://doi.org/10.1016/j.powtec.2021.07.082}
  {\bibfield  {journal} {\bibinfo  {journal} {Powder Technol.}\ }\textbf
  {\bibinfo {volume} {393}},\ \bibinfo {pages} {519--538} (\bibinfo {year}
  {2021})}\BibitemShut {NoStop}%
\bibitem [{\citenamefont {Gauthier}\ \emph {et~al.}(2021)\citenamefont
  {Gauthier}, \citenamefont {Pruvost}, \citenamefont {Gamache},\ and\
  \citenamefont {Colin}}]{gauthier2021new}%
  \BibitemOpen
  \bibfield  {author} {\bibinfo {author} {\bibfnamefont {A.}~\bibnamefont
  {Gauthier}}, \bibinfo {author} {\bibfnamefont {M.}~\bibnamefont {Pruvost}},
  \bibinfo {author} {\bibfnamefont {O.}~\bibnamefont {Gamache}},\ and\ \bibinfo
  {author} {\bibfnamefont {A.}~\bibnamefont {Colin}},\ }\bibfield  {title}
  {\enquote {\bibinfo {title} {A new pressure sensor array for normal stress
  measurement in complex fluids},}\ }\href {https://doi.org/10.1122/8.0000249}
  {\bibfield  {journal} {\bibinfo  {journal} {J. Rheol.}\ }\textbf {\bibinfo
  {volume} {65}},\ \bibinfo {pages} {583--594} (\bibinfo {year}
  {2021})}\BibitemShut {NoStop}%
\bibitem [{\citenamefont {Dally}\ and\ \citenamefont
  {Riley}(1991)}]{dally1991experimental}%
  \BibitemOpen
  \bibfield  {author} {\bibinfo {author} {\bibfnamefont {J.}~\bibnamefont
  {Dally}}\ and\ \bibinfo {author} {\bibfnamefont {W.}~\bibnamefont {Riley}},\
  }\href@noop {} {\emph {\bibinfo {title} {Experimental Stress Analysis}}},\
  McGraw-Hill series in mechanical engineering\ (\bibinfo  {publisher}
  {McGraw-Hill},\ \bibinfo {year} {1991})\BibitemShut {NoStop}%
\end{thebibliography}
\end{document}